\documentclass[12pt,aps,prb,preprint, fleqn]{revtex4}
\usepackage[dvips]{color}
\usepackage{amsmath}    % need for subequations
\usepackage{graphicx}   % for figures
\usepackage{wasysym}
\usepackage{eufrak}
\begin{document}
\definecolor{r}{rgb}{1,0,0}
\definecolor{g}{cmyk}{0,0,1,0}
\newcommand{\ab}{\v{a}} 
\newcommand{\ai}{\^{a}} 
\newcommand{\ib}{\^{\i}} 
\newcommand{\tb}{\c{t}} 
\newcommand{\st}{\c{s}}
\newcommand{\Ab}{\v{A}} 
\newcommand{\Ai}{\^{A}} 
\newcommand{\Ib}{\^{I}} 
\newcommand{\Tb}{\c{T}}
\newcommand{\St}{\c{S}}

\newcommand{\muv}{\boldsymbol{\mu}}
\newcommand{\mc}{{\mathcal M}}
\newcommand{\pc}{{\mathcal P}}
 \newcommand{\mv}{\boldsymbol{m}}
\newcommand{\pv}{\boldsymbol{p}}
\newcommand{\tv}{\boldsymbol{t}}
\def\msf{\hbox{{\sf M}}}
\def\msft{\boldsymbol{{\sf M}}}
\def\psf{\hbox{{\sf P}}}
\def\psft{\boldsymbol{{\sf P}}}
\def\Nsf{{\sf N}}
\def\Nsft{\boldsymbol{{\sf N}}}
\def\Tsf{\hbox{{\sf T}}}
\def\Tsft{\boldsymbol{{\sf T}}}
\def\Asf{\hbox{{\sf A}}}
\def\Asft{\boldsymbol{{\sf A}}}
\def\Bsf{\hbox{{\sf B}}}
\def\Bsft{\boldsymbol{{\sf B}}}
\def\Lsf{\hbox{{\sf L}}}
\def\Lsft{\boldsymbol{{\sf L}}}
\def\Ssf{\hbox{{\sf S}}}
\def\Ssft{\boldsymbol{{\sf S}}}
\def\Mtens{\bi{M}}
\def\msfsim{\boldsymbol{{\sf M}}_{\scriptstyle\rm(sym)}}
\newcommand{\mcsim}{ {\sf M}_{ {\scriptstyle \rm {(sym)} } i_1\dots i_n}}
\newcommand{\mcs}{ {\sf M}_{ {\scriptstyle \rm {(sym)} } i_1i_2i_3}}

\newcommand{\beqan}{\begin{eqnarray*}}
\newcommand{\eeqan}{\end{eqnarray*}}
\newcommand{\beqa}{\begin{eqnarray}}
\newcommand{\eeqa}{\end{eqnarray}}

 \newcommand{\suml}{\sum\limits}
 \newcommand{\sumd}{\suml_{\mathcal D}}
\newcommand{\intl}{\int\limits}
\newcommand{\ointl}{\oint\limits}
\newcommand{\rvec}{\boldsymbol{r}}
\newcommand{\xvec}{\boldsymbol{x}}
\newcommand{\xivec}{\boldsymbol{\xi}}
\newcommand{\Avec}{\boldsymbol{A}}
\newcommand{\Rvec}{\boldsymbol{R}}
\newcommand{\Evec}{\boldsymbol{E}}
\newcommand{\Bvec}{\boldsymbol{B}}
\newcommand{\Svec}{\boldsymbol{S}}
\newcommand{\avec}{\boldsymbol{a}}
\newcommand{\nablav}{\boldsymbol{\nabla}}
\newcommand{\nuvec}{\boldsymbol{\nu}}
\newcommand{\bvec}{\boldsymbol{\beta}}
\newcommand{\vvec}{\boldsymbolF}
\newcommand{\jvec}{\boldsymbol{J}}
\newcommand{\nvec}{\boldsymbol{n}}
\newcommand{\pvec}{\boldsymbol{p}}
\newcommand{\mvec}{\boldsymbol{m}}
\newcommand{\evec}{\boldsymbol{e}}
\newcommand{\eps}{\varepsilon}
\newcommand{\la}{\lambda}
\newcommand{\rad}{\mbox{\footnotesize rad}}
\newcommand{\scr}{\scriptstyle}
\newcommand{\latens}{\boldsymbol{\Lambda}}
\newcommand{\pitens}{\boldsymbol{\Pi}}
\newcommand{\cm}{{\cal M}}
\newcommand{\cp}{{\cal P}}
\newcommand{\beq}{\begin{equation}}
\newcommand{\eeq}{\end{equation}}
\newcommand{\ptens}{\boldsymbol{{\sf P}}}
\newcommand{\Ptens}{\boldsymbol{P}}
\newcommand{\Ttens}{\boldsymbol{{\sf T}}}
\newcommand{\Ntens}{\boldsymbol{{\sf N}}}
\newcommand{\Ncal}{\boldsymbol{{\cal N}}}
\newcommand{\Atens}{\boldsymbol{{\sf A}}}
\newcommand{\Btens}{\boldsymbol{{\sf B}}}
\newcommand{\dom}{\mathcal{D}}
\newcommand{\al}{\alpha}
\newcommand{\sym}{\scriptstyle \rm{(sym)}}
\newcommand{\Tcal}{\boldsymbol{{\mathcal T}}}
\newcommand{\Nmc}{{\mathcal N}}
\renewcommand{\d}{\partial}
\def\rmi{{\rm i}}
\def\rme{\hbox{\rm e}}
\def\rmd{\hbox{\rm d}}
\newcommand{\ct}{\mbox{\Huge{.}}}
\newcommand{\Laop}{\boldsymbol{\Lambda}}
\newcommand{\Ssfs}{{\scriptstyle \Ssft^{(n)}}}
\newcommand{\Lsfs}{{\scriptstyle \Lsft^{(n)}}}
\newcommand{\psfr}{\widetilde{\psf}}
\newcommand{\msfr}{\widetilde{\msf}}
\newcommand{\msftr}{\widetilde{\msft}}
\newcommand{\psftr}{\widetilde{\psft}}
\newcommand{\qdot}{\stackrel{\cdot\cdot\cdot\cdot}}
\newcommand{\tdot}{\stackrel{\cdot\cdot\cdot}}
\newcommand{\eref}{(\ref}
\newcommand{\bsy}{\boldsymbol}
\newcommand{\dotj}{\boldsymbol{\dot{J}}}
\newcommand{\psfs}{{\sf P}}
\newcommand{\msfs}{{\sf M}}
\newcommand{\Fvec}{\boldsymbol{F}}
\newcommand{\Qvec}{\boldsymbol{Q}}
\newcommand{\bmu}{\bsy{\mu}}
\newcommand{\bpi}{\bsy{\pi}}
\newcommand{\lasf}{{\sf \Lambda}}
\newcommand{\lasft}{\bsy{\sf \Lambda}}
\newcommand{\pisf}{{\sf \Pi}}
\newcommand{\pisft}{\bsy{\sf \Pi}}
\newcommand{\gamsf}{{\sf \Gamma}}
\newcommand{\gamsft}{\bsy{\sf \Gamma}}
\newcommand{\Scal}{\bsy{\mathcal S}}
\newcommand{\Tsfs}{{\sf T}}
\newcommand{\mfr}{\mathfrak{M}}
\newcommand{\nfr}{\mathfrak{N}}
\newcommand{\mifr}{\mathfrak{m}}
\newcommand{\mifrt}{\bsy{\mathfrak{m}}}
\newcommand{\mfrt}{\boldsymbol{\mfr}}
\newcommand{\pct}{\bsy{\pc}}
\newcommand{\mct}{\bsy{\mc}}
\title{Explanation notes on the  multipole expansions of the electromagnetic  field }
\author{C. Vrejoiu and R. Zus}
 \affiliation{University of Bucharest, Department of Physics, Bucharest, Romania} 
 \email{vrejoiu@fizica.unibuc.ro;roxana.zus@fizica.unibuc.ro}   %optional
\date{\today}

\begin{abstract}
Starting from Jefimenko's equations, we consider the multipole expansions of electric and magnetic fields for a confined system of charges and currents. We analyze and comment on the calculus of radiated power, on the consistent use of approximation criteria, on the invariance of physical results when changing the point of reference, as well as on the use of electric and magnetic moments described by symmetric and trace free tensors.
\end{abstract}

\maketitle

\section{Introduction}\label{sec:intro}
Despite the successful and long history of the electromagnetic field theory, there are several topics open to new theoretical and pedagogical contributions. One of them concerns the formalism of the multipole expansion of fields, in general, and of the radiated one, in particular. Another issue is related to the importance of Jefimenko's equations in the study of such problems.  Motivated by recent publications on the topic, we discuss some features of this type of problems, the present  paper being a revisited and completed version of a previous electronic preprint \cite{cvrz}. 
\par The multipole expansion of the electromagnetic field in Cartesian coordinates is exposed in electrodynamics textbooks, as the well-known Refs. \cite{Landau} and \cite{Jackson}.
 Ordinarily, these expansions are calculated only in the first  or second approximation, the higher-order terms being considered too complicated. As Jackson writes in his textbook, {\it the labor involved in manipulating terms in the expansion of the vector potential becomes increasingly prohibitive as the expansion is extended beyond the electric quadrupole terms} (see Ref. \cite{Jackson}, pp 415-416). 
 For this reason and due to the applicability only in the long-wavelength range, another treatment, based on the spherical tensors and on the solutions of Helmholtz equation is preferred. This alternative has also a larger domain of applications. Actually, starting from the results obtained employing this calculation technique, the reader can verify what effort is involved when returning to the multipole Cartesian moments which offer a higher physical transparency (see Ref. \cite{RV}). A relatively recent textbook \cite{Raab} and a paper \cite{Melo}, the last  related to the importance of Jefimenko's equations for expressing the electric and magnetic field when discussing the radiation theory, brought our attention on the necessity of some supplementary explanations. There are some prescriptions in the literature \cite{Thorne}, \cite{Damour}, \cite{cvcs} for calculating higher-order terms of the multipole series based on a simple algebraic formalism of tensorial analysis. One of the aims of the present paper is to show how one can hide, as much as possible, the higher-order tensors behind some vectors, reducing the calculation technique to the formalism of an ordinary vectorial algebra or analysis. 
 \par We start in section \ref{General} by shortly presenting the notation convention we use  by giving a general formalism for handling multipolar expansions in Cartesian coordinates.  In section \ref{Field expansion}, we derive the  electric and magnetic fields multipole expansions  without using the retarded potentials, while  in section \ref{STF} 
 we present some fundamental  ideas of expressing the multipole moments by symmetric trace free tensors. In section \ref{criterii} we discuss the approximation criteria stressing the significance of $d/\la$-criterion. 
  In section \ref{wave region}, we further discuss some features of the calculation for the radiated power, with an emphasis on the $4-th$ order approximation in $d/\la\,<\,1$. Section \ref{translation} presents the problem of the translational invariance of the fields expansions.
  Finally, in appendix \ref{sec:formulae}, we give the guidelines for the general tensorial calculus of the electric and magnetic moments and some general results in terms of symmetric trace free tensors. The last section is reserved for conclusions.     

\section{General Formalism}
\label{General}
Let us  write  Maxwell's  equations with a ``system free'' notation (a notation independent of the unit system):
\beqa\label{Maxwell}
\nablav\times\Bvec=\frac{\mu_0}{\al}\left(\jvec+\eps_0\frac{\d\Evec}{\d t}\right),\;\;\nablav\times\Evec=-\frac{1}{\al}\frac{\d\Bvec}{\d t},\;\;\;
\nablav\cdot\Bvec=0,\;\;\;\nablav\cdot\Evec=\frac{1}{\eps_0}\rho
\eeqa
where $\eps_0,\,\mu_0,\,\al$ are proportional  factors  depending on the system of units and satisfying the equation
\beqa\label{relc}
\frac{\al^2}{\eps_0\mu_0}=c^2.
\eeqa
$c$ is the vacuum light speed.
Maxwell equations written in SI units are obtained from equations \eref{Maxwell}) for  $\al=1$ and the SI values of $\eps_0,\;\mu_0$. For the Gauss system of units, $\al=c,\,\eps_0=1/4\pi,\,\mu_0=4\pi$.
\par The relations between fields and potentials are written as 
\beqa\label{AF}
\Bvec=\nablav\times\Avec,\;\;\Evec=-\nablav\Phi-\frac{1}{\al}\frac{\d\Avec}{\d t}.
\eeqa

Ordinarily, for  the field calculation  when the charge and the current distributions are known as functions of $\rvec$ and $t$, one firstly determines the retarded potentials: 
\beqa\label{Pret}
\Avec(\rvec,t)=\frac{\mu_0}{4\pi\al}\intl_{\dom}\frac{\jvec(\rvec',t-\frac{R}{c})}{R}\,\rmd^3x',\;\;\;
\Phi(\rvec,t)=\frac{1}{4\pi\eps_0}\intl_{\dom}\frac{\rho(\rvec',t-\frac{R}{c})}{R}\,\rmd^3x'.
\eeqa
The fields $\Evec,\,\Bvec$ are then derived from equations \eref{AF}).
The domain $\dom$ includes the supports of $\rho$ and $\jvec$, and $\Rvec=\rvec-\rvec'$. The solutions \eref{Pret}) are indeed electromagnetic potentials verifying the Lorenz condition 
\beqa\label{Lc}
\nablav\cdot\Avec+\frac{\eps_0\mu_0}{\al}\frac{\d\Phi}{\d t}=0.
\eeqa
This equation can be directly verified  using the continuity equation at the retarded time $\tau=t-R/c$:  
\beqa\label{econt}
\frac{\d\rho(\rvec',\tau)}{\d t}+\left[\nablav'\cdot\jvec(\rvec',t')\right]_{t'=\tau}=0
\eeqa 
and the relation between the retarded value of a spatial derivative of a function and the same spatial derivative of the retarded function: 
\beqa\label{dfret}
\d'_if(\rvec',\,\tau)
=\left[\d'_i\,f(\rvec',\,t')\right]_{t'=\tau}+\frac{R_i}{cR}\frac{\d}{\d t}f(\rvec',\tau).
\eeqa
The last equation should be well-known for each student from a class of electrodynamics since when writing the retarded potentials as particular solutions of the wave equation, it is necessary to verify the Lorenz condition. Only in this way one can be convinced that the retarded potentials are indeed electromagnetic potentials. This verification can be realized in a direct calculation, a good exercise for the student.
\par An alternative approach for the field calculation is given by Jefimenko's equations (see Ref. \cite{Jackson}, section 6.5). The fields are obtained directly as retarded solutions   of the  following wave equations which are consequences of  Maxwell's equations: 
\beqa\label{WEB}
\Delta\Bvec-\frac{1}{c^2}\frac{\d^2\Bvec}{\d t^2}=-\frac{\mu_0}{\al}\nablav\times\jvec,\;\;\;
\Delta\Evec-\frac{1}{c^2}\frac{\d^2\Evec}{\d t^2}=\frac{1}{\eps_0}\,\nablav\rho+\frac{\mu_0}{\al^2}\frac{\d\jvec}{\d t}.
\eeqa
These equations allow the retarded solutions 
\beqa\label{EBret}
\Bvec(\rvec,t)&=&\frac{\mu_0}{4\pi\al}\intl_{\dom}\frac{1}{R}\left[\nablav'\times\jvec(\rvec',t')\right]_{t'=\tau}\,\rmd^3x' , \nonumber\\
\Evec(\rvec,t)&=&-\frac{1}{4\pi\eps_0}\intl_{\dom}\frac{1}{R}\left[\nabla'\rho(\rvec',t')\right]_{t'=\tau}\rmd^3x'
-\frac{\mu_0}{4\pi\al^2}\intl_{\dom}\frac{1}{R}\frac{\d \jvec(\rvec',\tau)}{\d t}\,\rmd^3x'.
\eeqa
 Using equation \eref{dfret}), the following expressions of the solutions \eref{EBret}) can be obtained: 
\beqa\label{EBret'}
\Bvec(\rvec,t)&=&\frac{\mu_0}{4\pi\al}\intl_{\dom}\nablav\times\frac{\jvec(\rvec',\tau)}{R}\,\rmd^3x',\nonumber\\
\Evec(\rvec,t)&=&-\frac{1}{4\pi\eps_0}\intl_{\dom}\nablav\frac{\rho(\rvec',\tau)}{R}\,\rmd^3x'
-\frac{\mu_0}{4\pi\al^2}\intl_{\dom}\frac{\dot{\jvec}(\rvec',\tau)}{R}\,\rmd^3x',
\eeqa
where the dot signifies the time derivative. 
These last expressions can be obtained also introducing equations \eref{Pret}) in equations \eref{AF}) and inverting  the order of the derivatives  and integrals. Considered as solutions of equations \eref{WEB}), the expressions \eref{EBret'}) are known as {\it Jefimenko's equations}. We think it is important to show that  sometimes it is convenient to perform a  calculation directly on the fields represented by equations \eref{EBret'}) instead of  calculating firstly the potentials and after that, applying equations \eref{AF}). In fact, in the present work we benefit from  this circumstance. We owe our inspiration to  Ref.\cite{Melo}. 
\section{Multipole expansions of fields} 
 \label{Field expansion}
Let us consider the field $\Bvec(\rvec,t)$, given by equation \eref{EBret'}), in a point from the exterior of the domain $\dom$ which includes the supports of $\rho$ and $\jvec$: 
\beqa\label{B1}
\!\!\!\!\!\!\!\!\!\!\!\!\!\Bvec(\rvec,t)=\frac{\mu_0}{4\pi\al}\,\evec_i\eps_{ijk}\,\intl_{\dom}\,\d_j\frac{J_k(\rvec',\tau)}{R}\,\rmd^3x'
=\frac{\mu_0}{4\pi\al}\evec_i\eps_{ijk}\intl_{\dom}\left[\d_j\frac{J_k(\rvec',t-\frac{\vert\rvec-\xivec\vert}{c})}{\vert\rvec-\xivec\vert}\right]_{\xivec=\rvec'}\rmd^3x'.
\eeqa
$\evec_i$ are the unit vectors of the Cartesian axes. 
Writing the Taylor series  about $\xivec=0$ of the integrand, we can finally write 
\beqa\label{B2}
 \Bvec(\rvec,t)=\frac{\mu_0}{4\pi\al}\,\evec_i\eps_{ijk}\,\intl_{\dom}\rmd^3x'
 \suml_{n\ge 0}\frac{(-1)^n}{n!}\,x'_{i_1}\dots x'_{i_n}\,\d_j\d_{i_1}\dots \d_{i_n}\frac{J_k(\rvec',\tau_0)}{r} 
\eeqa
where $\tau_0=t-r/c$. Here, we can apply a formula for the multiple derivative of a function of the type $F(\tau_0)/r$: 
\beqa\label{deriv1}
\d_{i_1}\dots\d_{i_n}\frac{F(\tau_0)}{r}=\suml^n_{l=0}\frac{1}{c^{n-l}r^{l+1}}C^{(n,\,l)}_{i_1\dots i_n}\frac{\d^{n-l}F(\tau_0)}{\d t^{n-l}}.
\eeqa
The coefficients are symmetric in $i_1,\dots,i_n$ and can be expressed as 
\beqa\label{deriv2}
C^{(n,\,l)}_{i_1\dots i_n}=\suml^{[\frac{n}{2}]}_{k=0}D^{(n,\,l)}_k\,\delta_{\{i_1\,i_2}\dots \delta_{i_{2k-1}\,i_{2k}}
\nu_{2k+1}\dots\nu_{i_n\}}.
\eeqa
In the last equation, $[\beta]$ is the integer part of $\beta$ and $\nuvec=\rvec/r$. By 
$A_{\{i_1\dots i_n\}}$ 
we understand the sum over all the  permutations of the symbols $i_q$ giving distinct terms. For the objective of the present paper ($n\le 3$), the coefficients $D$ in equation \eref{deriv2}) can be calculated directly from the successive 
derivative operations but, generally, one can establish reccurence relations. 
We can write now the expansion \eref{B2}) as 
\beqa\label{B3}
\!\!\!\!\!\!\!\!\!\!\!\!\!\!\!\Bvec(\rvec,t)= \frac{\mu_0}{4\pi\al}\evec_i\eps_{ijk} \suml_{n\ge 0}\frac{(-1)^n}{n!}\,
\suml^{n+1}_{l=0}\frac{1}{c^{n+1-l}\,r^{l+1}}\,C^{(n+1,\,l)}_{j\,i_1\dots i_n}
 \intl_{\dom}\rmd^3x'\,x'_{i_1}\dots x'_{i_n}\frac{\d^{n+1-l}J_k(\rvec',\tau_0)}{\d t^{n+1-l}}.
\eeqa
Next, we admit that in the previous equation one can invert the derivation with the integral operation on the domain  $\dom$ and we have  
\beqa\label{B1-0}
\Bvec(\rvec,t)= \frac{\mu_0}{4\pi\al}\,\evec_i\eps_{ijk} \suml_{n\ge 0}\frac{(-1)^n}{n!}\,
\suml^{n+1}_{l=0}\frac{1}{c^{n+1-l}\,r^{l+1}}\,C^{(n+1,\,l)}_{j\,i_1\dots i_n}
\frac{\d^{n+1-l}}{\d t^{n+1-l}}\mfr_{i_1\dots i_n\,k}(\tau_0) ,
\eeqa
where it is introduced the $n-th$ order magnetic moment $\mfrt^{(n)}$ with the Cartesian components: 
\beqa\label{Mn0}
\mfr_{i_1\dots i_n}(t)=\intl_{\dom}\,x'_{i_1}\dots x'_{i_{n-1}}J_{i_n}(t)\,\rmd^3x'.
\eeqa
From equation \eref{EBret'}), performing a similar calculation  as in the magnetic field case, one obtains the multipole expansion of the electric field: 
\beqa\label{E1-0}
\Evec(\rvec,t)&=&-\frac{1}{4\pi\eps_0}\evec_i\suml_{n\ge 0}\frac{(-1)^n}{n!}\suml^{n+1}_{l=0}\frac{1}{c^{n+1-l}r^{l+1}}\,C^{(n+1,\,l)}_{i\,i_1\dots i_n}
\frac{\d^{n+1-l}}{\d t^{n+1-l}}\,\psfs_{i_1\dots i_n}(\tau_0)\nonumber\\
&&-\frac{\mu_0}{4\pi\al^2}\evec_i\suml_{n\ge 0}\frac{(-1)^n}{n!}
\suml^n_{l=0}\frac{1}{c^{n-l}r^{l+1}}\,C^{(n,\,l)}_{i_1\dots i_n}
\frac{\d^{n+1-l}}{\d t^{n+1-l}}\mfr_{i_1\dots i_n i}(\tau_0).
\eeqa
$\psfs_{i_1\dots i_n}$ are the Cartesian components of the electric moment of order $n$, $\psft^{(n)}$: 
\beqa\label{Pn}
\psfs_{i_1\dots i_n}(t)=\intl_{\dom}x'_{i_1}\dots x'_{i_n}\,\rho(\rvec',t)\,\rmd^3x'.
\eeqa
 We point out that the  expressions from equations \eref{B1-0}) and \eref{E1-0}) are obtained without using commuting properties of the different operations except for the integration and the Taylor series expansion and, also, for the time derivation and the spatial integration. 
Of course, all the necessary convergence properties are supposed satisfied. In equation \eref{B1-0}) and \eref{E1-0})  one recognizes the multiple spatial derivative of  $ \mfr_{i_1\dots i_n\,k}(\tau_0)/r$ and  $\psfs_{i_1\dots i_n}(\tau_0)/r$ as, for example, 
\beqa\label{mder}
\suml^{n+1}_{l=0}\frac{1}{c^{n+1-l}r^{l+1}}\,C^{(n+1,\,l)}_{j\,i_1\dots i_n}\frac{\d^{n+1-l}}{\d t^{n+1-l}}\,\mfr_{i_1\dots i_n\,k}(\tau_0)
=\d_j\,\d_{i_1}\dots \d_{i_n}\big[\frac{1}{r}\mfr_{i_1\dots i_n\,k}(\tau_0)\big].
\eeqa
The fields $\Bvec$ and $\Evec$ can be expressed as 
\beqa\label{B1-2}
\Bvec(\rvec,t)
&=&\frac{\mu_0}{4\pi\al}\evec_i\,\eps_{ijk}\suml_{n\ge 0} \frac{(-1)^n}{n!}\,\d_j\,\d_{i_1}\dots\,\d_{i_n}\frac{\mfr_{i_1\dots i_n\,k}(\tau_0)}{r}\nonumber\\
&=&\frac{\mu_0}{4\pi\al}\nablav\times\,\suml_{n\ge 0}\frac{(-1)^n}{n!}\nablav^n\vert\vert\frac{\mfrt^{(n+1)}(\tau_0)}{r} 
\eeqa
and  
\beqa\label{E1-2}
\!\!\!\!\!\Evec(\rvec,t)&=&-\frac{1}{4\pi\eps_0}\,\evec_i\suml_{n\ge 0}\frac{(-1)^n}{n!}\,\d_i\,\d_{i_1}\dots\d_{i_n}\frac{\psfs_{i_1\dots i_n}(\tau_0)}{r}-
\frac{\mu_0}{4\pi\al^2}\,\evec_i\suml_{n\ge 0}\frac{(-1)^n}{n!}\,\d_{i_1}\dots\,\d_{i_n}\frac{\dot{\mfr}_{i_1\dots i_n\,i}}{r}\nonumber\\
\!\!\!\!\!&=&-\frac{1}{4\pi\eps_0}\nablav\,\suml_{n\ge 0}\frac{(-1)^n}{n!}\,\nablav^n\vert\vert\frac{\psft^{(n)}(\tau_0)}{r}
-\frac{\mu_0}{4\pi\al^2}\suml_{n\ge 0}\frac{(-1)^n}{n!}\,\nablav^n\vert\vert\frac{\dot{\mfrt}^{(n+1)}}{r},
\eeqa
where the following notation is introduced for the tensor contractions: 
\beqa\label{contr}
 ({\Atens}^{(n)}||{\Btens}^{(m)})_{i_1 \cdots i_{|n-m|}}
=\left\{\begin{array}{ll}
A_{i_1 \cdots i_{n-m}j_1 \cdots j_m}B_{j_1 \cdots j_m} & ,\; n>m\\
A_{j_1 \cdots j_n}B_{j_1 \cdots j_n} & ,\; n=m\\
A_{j_1 \cdots j_n}B_{j_1 \cdots j_n i_1 \cdots i_{m-n}} & ,\; n<m
\end{array} \right..
\eeqa
 We recognize, comparing these equations with equations \eref{Pret}), the multipole expansions of the potentials:
\beqa\label{Aexp}
\!\!\!\!\!\!\!\!\!\!\!\!\!\!\!\!\!&~&\Avec(\rvec,t)=\frac{\mu_0}{4\pi\al}\,\suml_{n\ge 0}\frac{(-1)^n}{n!}\nablav^n\vert\vert\frac{\mfrt^{(n+1)}(\tau_0)}{r},\;\;\;
\Phi(\rvec,t)=\frac{1}{4\pi\eps_0}\suml_{n\ge 0}\frac{(-1)^n}{n!}\nablav^n\vert\vert\frac{\psft^{(n)}(\tau_0)}{r}. 
\eeqa

Let us define 
\beqa\label{ak}
a_k(\rvec,t;\zeta,n)=\zeta_{i_1\dots i_n}\intl_{\dom}x'_{i_1}\dots x'_{i_n}\,J_k(\rvec',t)\,\rmd^3x'= \zeta_{i_1\dots i_n}\mfr_{i_1\dots i_n\,k}(t),
\eeqa
with $\zeta_{i_1\dots i_n}$ symmetric in the $n$ indices.
 Considering the following consequence of the continuity equation
\beqan
J_k(\rvec',\tau_0)=\nablav'\cdot\big[x'_k\jvec(\rvec',\tau_0)\big]+x'_k\frac{\d\rho(\rvec',\tau_0)}{\d t},
\eeqan
we can write
\beqan
a_k(\rvec,\tau_0;\zeta,n)
=-\zeta_{i_1\dots i_n}\intl_{\dom}x'_k\,\jvec(\rvec',\tau_0)\cdot\nablav'(x'_{i_1}\dots x'_{i_n})\,\rmd^3x'
+\zeta_{i_1\dots i_n}\dot{\psfs}_{i_1\dots i_n\,k}(\tau_0).
\eeqan
Partial integrations  and   vanishing surface integrals are employed for obtaining the last result.
 Further, because of the symmetry of $\zeta$, we get
 \beqan
 a_k(\rvec,\tau_0;\zeta,n)&-&\zeta_{i_1\dots i_n}\dot{\psfs}_{i_1\dots i_n\,k}(\tau_0)
=-n\,\zeta_{i_1\dots i_n}\intl_{\dom}x'_{i_1}\dots x'_{i_{n-1}}x'_kJ_{i_n}(\rvec',\tau_0)\,\rmd^3x'\\
 &=&-n\,\zeta_{i_1\dots i_n}\intl_{\dom}x'_{i_1}\dots x'_{i_{n-1}}\big[x'_kJ_{i_n}(\rvec',\tau_0)
-x'_{i_n}J_k(\rvec',\tau_0)\big]\,\rmd^3x'-n\, a_k(\rvec,\tau_0;\zeta,n)\\
&=&-n\,\zeta_{i_1\dots i_n}\eps_{ki_nq}\intl_{\dom}x'_{i_1}\dots x'_{i_{n-1}}
\big[\rvec'\times\jvec(\rvec',\tau_0)\big]_q\,\rmd^3x'
-n\, a_k(\rvec,\tau_0;\zeta,n).
 \eeqan
Let us introduce the standard magnetic moment of order $n$, $\msft^{(n)}$, by its Cartesian components \cite{Castell} \beqa\label{Mn}
\msfs_{i_1\dots i_n}(t)
=\frac{n}{(n+1)\al}\intl_{\dom}x'_{i_1}\dots x'_{i_{n-1}}
\big[\rvec'\times\jvec(\rvec',t)\big]_{i_n}\,\rmd^3x'. 
 \eeqa
 We can write 
  \beqa\label{1}
 \!\!\!\!\!\!\!\!\!\!\!\!\!\!\!\!a_k(\rvec,\tau_0;\zeta,n)\equiv \zeta_{i_1\dots i_n}\mfr_{i_1\dots i_n\,k}(\tau_0)=\zeta_{i_1\dots i_n}\left(-\al\,\eps_{ki_nq}\msfs_{i_1\dots i_{n-1}\,q}(\tau_0)
 +\frac{1}{n+1}\,\dot{\psfs}_{i_1\dots i_n\,k}(\tau_0)\right).
\eeqa
This result inserted in  equation \eref{B1-0}) gives  
\beqa\label{dB1}
\Bvec(\rvec,t)&=&\frac{\mu_0}{4\pi}\evec_i\eps_{ijk}\suml_{n\ge 1}\frac{(-1)^{n-1}}{n!}\,\eps_{ki_n q}\suml^{n+1}_{l=0}\frac{1}{c^{n+1-l}\,r^{l+1}}
 C^{(n+1,\,l)}_{j\,i_1\dots i_n}\frac{\d^{n+1-l}}{\d t^{n+1-l}}\,\msfs_{i_1\dots i_{n-1}\,q}(\tau_0)\nonumber\\
 &&+ \frac{\mu_0}{4\pi\al}\evec_i\eps_{ijk}\suml_{n\ge 0}\frac{(-1)^n}{(n+1)!}
\suml^{n+1}_{l=0}\frac{1}{c^{n+1-l}r^{l+1}}\,C^{(n+1,\,l)}_{j\,i_1\dots i_n}\,\frac{\d^{n+1-l}}{\d t^{n+1-l}}\,\dot{\psfs}_{i_1\dots i_n\,k}(\tau_0).
\eeqa
With the same calculation technique, one obtains the multipole expansion of the electric field: 
\beqa\label{dE1}
\Evec(\rvec,t)&=&-\frac{1}{4\pi\eps_0}\evec_i\suml_{n\ge 0}\frac{(-1)^n}{n!}\suml^{n+1}_{l=0}\frac{1}{c^{n+1-l}r^{l+1}}\,C^{(n+1,\,l)}_{i\,i_1\dots i_n}
\frac{\d^{n+1-l}}{\d t^{n+1-l}}\,\psfs_{i_1\dots i_n}(\tau_0)\nonumber\\
&&-\frac{\mu_0}{4\pi\al^2}\evec_i\suml_{n\ge 0}\frac{(-1)^n}{(n+1)!}\suml^n_{l=0}\frac{1}{c^{n-l}r^{l+1}}\,C^{(n,\,l)}_{i_1\dots i_n}\frac{\d^{n-l}}{\d t^{n-l}}\,\ddot{\psfs}_{i_1\dots i_n\,i}(\tau_0)\nonumber\\
&&+\frac{\mu_0}{4\pi\al}\evec_i\suml_{n\ge 1}\frac{(-1)^n}{n!}\eps_{ii_nq}
\suml^n_{l=0}\frac{1}{c^{n-l}r^{l+1}}\,C^{(n,\,l)}_{i_1\dots i_n}\,\frac{\d^{n-l}}{\d t^{n-l}}\dot{\msfs}_{i_1\dots i_{n-1}q}(\tau_0).
\eeqa
Processing the expressions from equations \eref{dB1}) and \eref{dE1}) by the same procedure applied for obtaining equations \eref{B1-2}) and \eref{E1-2}), we obtain 
 \beqa\label{dB2}
\!\!\!\!\!\!\!\!\!\!\!\!\Bvec(\rvec,t)=
\frac{\mu_0}{4\pi}\nablav\times\left[\suml_{n\ge 1}\frac{(-1)^{n-1}}{n!}\nablav\times\left(\nablav^{n-1}\vert\vert\frac{\msft^{(n)}(\tau_0)}{r}\right)+\frac{1}{\al}\suml_{n\ge 0}\frac{(-1)^n}{(n+1)!}\,\nablav^n\vert\vert\frac{\dot{\psft}^{(n+1)}}{r}\right]
\eeqa
\par 
and 
\beqa\label{dE2}
\Evec(\rvec,t)&=&
-\frac{1}{4\pi\eps_0}\left[\nablav\suml_{n\ge 0}\frac{(-1)^n}{n!}\nablav^n\vert\vert\frac{\psft^{(n)}(\tau_0)}{r}
+\frac{1}{c^2}\suml_{n\ge 0}\frac{(-1)^n}{(n+1)!}\nablav^n\vert\vert\frac{\ddot{\psft}^{(n+1)}(\tau_0)}{r}\right.\nonumber\\
&+&\left. \frac{\al}{c^2}  \suml_{n\ge  1}\frac{(-1)^{n-1}}{n!}\nablav\times\left(\nablav^{n-1}\vert\vert\frac{\dot{\msft}^{(n)}(\tau_0)}{r}\right)\right].
\eeqa
From equations  \eref{AF}), \eref{dB2}) and \eref{dE2}) one can immediately identify the multipole expansions of the potentials $\Avec$ and $\Phi$. 
\par The expansions from equations \eref{dB2}) and \eref{dE2}) can be easily processed by a vectorial calculus  introducing the following vectors:
\beqa\label{pcal}
\pct(\rvec,t;\xivec,n)&=&\evec_i\xi_{i_1}\dots\xi_{i_{n-1}}\,\frac{\psf_{i_1\dots i_{n-1}\,i}(\tau_0)}{r}=\xivec^{n-1}\vert\vert\frac{\psft^{(n)}(\tau_0)}{r}
\eeqa
and
 \beqa\label{mcal}
 \mct(\rvec,t;\xivec,n)&=&\evec_i\xi_{i_1}\dots\xi_{i_{n-1}}\,\frac{\msf_{i_1\dots i_{n-1}\,i}(\tau_0)}{r}=\xivec^{n-1}\vert\vert\frac{\msft^{(n)}(\tau_0)}{r}.
 \eeqa
 Consequently, the expressions for the magnetic and electric field become
  \beqa\label{dB3}
\!\!\!\!\!\!\!\!\!\!\!\!\Bvec(\rvec,t)=\frac{\mu_0}{4\pi}\nablav\times\,\left[\suml_{n\ge 1}\frac{(-1)^{n-1}}{n!}\nablav\times\mct(\rvec,t;\nablav,n)+\frac{1}{\al}\suml_{n\ge 0}\frac{(-1)^n}{(n+1)!}\dot{\pct}(\rvec,t;\nablav,n+1)\right]
\eeqa
and 
\beqa\label{dE3}
\!\!\!\!\!\!\!\!\!\!\!\!\Evec(\rvec,t)&=&-\frac{1}{4\pi\eps_0}\left[\nablav\suml_{n\ge 0}\frac{(-1)^n}{n!}\big(\nablav\cdot\pct(\rvec,t;\nablav,n)\big)
+\frac{1}{c^2}\suml_{n\ge 0}\frac{(-1)^n}{(n+1)!}\ddot{\pct}(\rvec,t;\nablav,n+1)\right.\nonumber\\
\!\!\!\!\!\!\!\!\!\!\!\!&&+\left.\frac{\al}{c^2}\suml_{n\ge 1}\frac{(-1)^{n-1}}{n!}\,\nablav\times\dot{\mct}(\rvec,t;\nablav,n)\right].
\eeqa
\section{Expressing the fields by symmetric trace-free tensors } 
\label{STF}
The general terms from the field expansions, given by equations \eref{dB1}) and \eref{dE1}) or \eref{dB2}) and \eref{dE2}), are  apparently too complicated  and they imply  cumbersome calculations when the fields $\Evec$ and $\Bvec$ are introduced in some physical interesting expressions as, for example, the radiation intensity. A considerable simplification of these calculations is obtained introducing the {\it irreducible} electric and magnetic moments defined as symmetric trace-free ({\bf ``STF''}) Cartesian tensors. Let us consider a $n-th$ order tensor $\Tsft^{(n)}$ 
and the corresponding projections $\Scal(\Tsft^{(n)})$ and $\Tcal(\Tsft^{(n)})$ on the subspaces of symmetric  and {\bf STF} tensors, respectively. For the  electric moment $\psft^{(n)}$, we have  a symmetric tensor and one has only to establish their trace free projection. Let us consider the simplest case of the quadrupolar moment $\psft^{(2)}$. Writing the components $\psfs_{ij}$ as
\beqan
\psfs_{ij}={\sf \Pi}_{ij}+\Lambda\,\delta_{ij},
\eeqan
there is a unique value of the parameter $\Lambda$ such that $\pitens^{(2)}=\Tcal(\psft^{(2)})$. For $\Lambda=\psfs_{qq}/3$, 
\beqa\label{67}
\pisf_{ij}=\psfs_{ij}-\frac{1}{3}\,\psfs_{qq}\,\delta_{ij}=\intl_{\dom}\big(x_ix_j-\frac{1}{3}\,r^2\,\delta_{ij}\big)
\,\rho\,\rmd^3x.
\eeqa
The {\bf STF} projection of $\psft^{(3)}$ can be calculated searching the first order tensor $\lasft^{(1)}$ such that $\pisft^{(3)}=\Tcal(\psft^{(3)})$ is given by the components 
\beqa\label{68}
\pisf_{ijk}=\psfs_{ijk}-\,\delta_{\{ij}\lasf_{k\}}.
\eeqa
From the condition of vanishing traces of the tensor $\pisft^{(3)}$, one easily obtains: 
\beqa\label{69}
\lasf_i=\frac{1}{5}\,\psfs_{qqi}=\frac{1}{5}\intl_{\dom}r^2\,x_i\,\rho\,\rmd^3x.
\eeqa
Concerning the magnetic quadrupole moment $\msft^{(2)}$, we have a simple procedure for obtaining the {\bf STF} projection. Let us write the identity
\beqan
\msfs_{ij}=\frac{1}{2}(\msfs_{ij}+\msfs_{ji})+\frac{1}{2}(\msfs_{ij}-\msfs_{ji}),
\eeqan
where the first bracket represents the symmetric part of this tensor, and the second, the antisymmetric one. The symmetric part is, for this case ($n=2$), a {\bf STF} tensor $\gamsft^{(2)}=\Tcal(\msft^{(2)})$. Therefore, 
\beqa\label{70}
\msfs_{ij}=\gamsf_{ij}+\frac{1}{2}\,\eps_{ijk}\,\Nsf_k,
\eeqa 
where 
\beqa\label{71}
\Nsf_k&=&\eps_{kij}\,\msfs_{ij}=\frac{2}{3\al}\intl_{\dom}\big[\rvec\times(\rvec\times\jvec)\big]_k\,\rmd^3x
=\frac{2}{3\al}\intl_{\dom}\big[(\rvec\cdot\jvec)\,x_k-r^2\,J_k\big]\,\rmd^3x.
\eeqa
We consider now the effect of the substitution 
\beqa\label{subst}
\psft^{(2)}\longrightarrow \pisft^{(2)}:\;\;\psfs_{ij}\longrightarrow \psfs_{ij}-\Lambda\,\delta_{ij}
\eeqa
in the multipole expansion of the field. From equation \eref{pcal}) we obtain the transformation of $\pct(\rvec,t;\nablav,\,2)$ by the substitution \eref{subst}): 
\beqa\label{trpc}
\pct(\rvec,t;\nablav,\,2)\stackrel{\psft^{(2)}\to \pisft^{(2)}}{\longrightarrow}\;\pct(\rvec,t;\nablav,\,2)-\nablav\,\frac{\Lambda(\tau_0)}{r}.
\eeqa
It is easy to see from equation \eref{dB3}) that $\Bvec$ is invariant to the substitution \eref{trpc}). In the expression \eref{dE3}) of the electric field $\Evec$, the combination of the electric quadripolar terms from the first two sums gets
\beqan
\frac{1}{8\pi\eps_0}\,\nablav\left[\Delta\,\frac{\Lambda(\tau_0)}{r}-\frac{1}{c^2}\,\frac{\ddot{\Lambda}(\tau_0)}{r}\right]=0
\eeqan
since, for $r\ne 0$, $\Lambda(\tau_0)/r$ is a solution of the homogeneous wave equation. Therefore, the field $\Evec$ is also invariant to the substitution \eref{trpc}).
\par The circumstances change dramatically for $n\ge 3$ in the cases of electric moments and for $n\ge 2$ in the magnetic ones. As  we will see in the following, the electromagnetic field is not invariant under the  substitutions 
$\psft^{(3)}\to\pisft^{(3)}$ and $\msft^{(2)}\to\gamsft^{(2)}$. The general case, i.e. for $n> 3$ for electric moments and $n > 2$ for the magnetic ones, is developed  in earlier works (see Ref.  \cite{cvcs} and the literature cited in). We insert in the appendix some general results from these works, without demonstrations. 
\par Let us consider the substitution 
\beqa\label{trp3}
\ptens^{(3)}\to \pisft^{(3)}:\;\;\psfs_{ijk}\to \psfs_{ijk}-\Lambda_{\{i}\,\delta_{jk\}},
\eeqa 
in the fields expressions. In this case, 
\beqa\label{trp3v}
\pct(\rvec,t;\nablav,3)&\to&
\pct(\rvec,t;\nablav,3)-2\nablav\left[\nablav\cdot\frac{\bsy{\Lambda}}{r}\right]-\frac{1}{c^2}\frac{\ddot{\bsy{\Lambda}}(\tau_0)}{r}\nonumber\\
&=&\pct(\rvec,t;\nablav,3)-2\,\nablav\times\big(\nablav\times\frac{\bsy{\Lambda}}{r}\big)-\frac{3}{c^2}\,\frac{\ddot{\bsy{\Lambda}}(\tau_0)}{r},
\eeqa
where $\bsy{\Lambda}=\evec_i\Lambda_i$.
Performing the substitution \eref{trp3v}) in equation \eref{dB3}), we obtain the following transformation of $\Bvec$:
\beqa\label{trB3}
\Bvec(\rvec,t)\rightarrow \Bvec(\rvec,t)-\frac{\mu_0}{24\pi\al c^2}\,\nablav\times\frac{\ddot{\bsy{\Lambda}}(\tau_0)}{r}.
\eeqa
The same substitution in equation \eref{dE3}) gives 
\beqa\label{trE3}
\Evec(\rvec,t)&\rightarrow& \Evec(\rvec,t)
-\frac{1}{24\pi\eps_0c^2}\,\nablav\big[\nablav\cdot\frac{\ddot{\bsy{\Lambda}}(\tau_0)}{r}\big]+\frac{\mu_0}{24\pi\al^2c^2}\,\frac{\qdot{\bsy{\Lambda}}(\tau_0)}{r}.
\eeqa
Both, the modifications of the magnetic and electric field given in equations \eref{trB3}) and \eref{trE3}), respectively, can be compensated by the following transformation of the electric dipolar moment: 
\beqa\label{modelp}
\pvec\rightarrow \pvec+\frac{1}{6c^2}\,\ddot{\bsy{\Lambda}}.
\eeqa
Let be the substitution 
\beqa\label{trm2}
\msft^{(2)}\rightarrow \gamsft^{(2)}:\;\;\msfs_{ij}\rightarrow \msf_{ij}-\frac{1}{2}\eps_{ijk}\,\Nsf_k,
\eeqa 
corresponding to equation \eref{70}). Then,
\beqan
\mct(\rvec,t;\nablav,2)\rightarrow \mct(\rvec,t;\nablav,2)-\frac{1}{2}\,\nablav\times\frac{\Nsft(\tau_0)}{r},
\eeqan
where $\Nsft=\evec_i\Nsf_i$, and 
\beqa\label{trmcal}
\nablav\times\mct(\rvec,t;\nablav,2)\rightarrow\nablav\times\mct(\rvec,t;\nablav,2)
+\frac{1}{2}\nablav\big[\nablav\cdot\frac{\Nsft(\tau_0)}{r}\big]-\frac{1}{2c^2}\,\frac{\ddot{\Nsft}(\tau_0)}{r}.
\eeqa
Performing the substitution \eref{trmcal}) in equations \eref{dB3}) and \eref{dE3}), we obtain the transformations:
\beqa\label{trB3a}
\Bvec(\rvec,t)\rightarrow \Bvec(\rvec,t)+\frac{\mu_0}{16\pi c^2}\,\nablav\times\frac{\ddot{\Nsft}(\tau_0)}{r}.
\eeqa
and 
\beqa\label{trE3a}
\Evec(\rvec,t)\rightarrow \Evec(\rvec,t)
+\frac{\mu_0}{16\pi\al}\nablav\big(\nablav\cdot\frac{\dot{\Nsft}(\tau_0)}{r}\big)-\frac{\mu_0}{16\pi\al c^2}\frac{\tdot{\Nsft}(\tau_0)}{r}.
\eeqa
The changes  of $\Bvec$ and $\Evec$ described by equations \eref{trB3a}) and \eref{trE3a}) are compensated by the transformation of electric dipolar moment, too: 
\beqa\label{modelpa}
\pvec\rightarrow \pvec-\frac{\al}{4c^2}\dot{\Nsft}.
\eeqa
Finally, substituting in the field expressions the momenta $\psft^{(3)}$ and $\msft^{(2)}$ by their {\bf STF}- projections $\pisft^{(3)}$ and $\gamsft^{(2)}$, respectively, the corresponding changes of the fields can be compensated by the transformation 
\beqa\label{tfinp}
\pvec\rightarrow \widetilde{\pvec}=\pvec-\frac{1}{c^2}\left(\frac{\al}{4}\,\dot{\Nsft}-\frac{1}{6}\,\ddot{\lasft}\right)=
\pvec\,-\,\frac{1}{c^2}\,\dot{\bsy{t}},
\eeqa
where by the vector $\bsy{t}$ we understand 
\beqa\label{deft}
\bsy{t}=\frac{\al}{4}\,\dot{\Nsft}-\frac{1}{6}\,\ddot{\lasft}=\frac{1}{10}\intl_{\dom}\big(\,(\rvec\cdot\jvec)\,\rvec\,-\,2\,r^2\,\jvec\big)\,\rmd^3x.
\eeqa
The last expression is obtained from equations \eref{69}) and \eref{71}) applying the continuity equation together with an operation of partial integration. This is the so-called {\it electric toroidal dipolar moment}, its presence in equation \eref{tfinp}) leading to physical effects similar to the ones produced by an electric dipole moment. This fact was observed firstly in Ref. \cite{Zeld}(see also Refs. \cite{Dubovik-FEC}, \cite{Dubovik-rep},\cite{Bellotti})).

\section{Approximation criteria} 
\label{criterii}
Generally, one can operate with the spectral decomposition of the charge and current distributions: 
\beqa\label{spectral}
\rho(\rvec,t)=\intl^\infty_{-\infty}\rho_\omega(\rvec)\,\rme^{-\rmi\omega t}\,\rmd\omega,\;\;\;\;\;
\jvec(\rvec,t)=\intl^\infty_{-\infty}\jvec_\omega(\rvec)\,\rme^{-\rmi\omega t}\,\rmd\omega.
\eeqa
Correspondingly, one can introduce the spectral decomposition of the multipole moments: 
\beqa\label{sppm}
\psft^{(n)}(t)=\intl^\infty_{-\infty}\,\psft_\omega^{(n)}\,\rme^{-\rmi\omega t}\,\rmd\omega,\;\;\;\;\;
\msft^{(n)}(t)=\intl^\infty_{-\infty}\,\msft_\omega^{(n)}\,\rme^{-\rmi\omega t}\,\rmd\omega.
\eeqa
The Fourier components are given by 
\beqa\label{FPM}
\!\!\!\!\!\!\!\psfs_{(\omega)i_1\dots  i_n}= \intl_{\dom}x'_{i_1}\dots x'_{i_n}\,\rho_\omega(\rvec')\,\rmd^3x',\;\;
\msfs_{(\omega)i_1\dots  i_n}= \intl_{\dom}x'_{i_1}\dots x'_{i_{n-1}}\big(\rvec'\times\jvec_\omega(\rvec')\big)_{i_n}\,\rmd^3x'. 
\eeqa
Let us denote by $d$ the linear dimension of the domain $\dom$. For estimating the order of magnitude of the different terms from the multipolar expansions of the fields, we notice the following relations: 
\beqa\label{ord-d}
\psft^{(n)}\,\sim\,d^n,\;\;\mfrt^{(n)}\,\sim\,d^{n+1},\;\;\msft^{(n)}\,\sim\,d^{n+1}.
\eeqa
Considering the Fourier components in equations \eref{B1-0}) and \eref{E1-0}), we can estimate the contributions of the parameters $d,\,\la=2\pi c/\omega$ and $r$  to the orders of magnitude of the different terms from the series expansions. For example, the estimation of the order of magnitude  in equation \eref{B1-0}) leads to the relations:
\beqan
S^{(n)}_{j\,i_1\dots i_n\,k}(\omega)&=&\suml^{n+1}_{l=0}\frac{1}{c^{n+1-l}\,r^{l+1}}\,C^{(n+1,\,l)}_{j\,i_1\dots i_n}
\,\frac{\d^{n+1-l}}{\d t^{n+1-l}}\left(\mfr_{(\omega)i_1\dots i_{n-1}\,k}\rme^{-\rmi \omega t+\rmi\omega r/c}\right)\\
&\sim& \frac{1}{r}\suml^{n+1}_{l=0}\frac{\omega^{n+1-l}\,d^{n+1}}{c^{n+1-l}\,r^l}\sim\frac{1}{r}\suml^{n+1}_{l=0} \frac{d^{n+1}}{\la^{n+1-l}\,r^l}
\eeqan
that is 
\beqa\label{ordF}
S^{(n)}_{j\,i_1\dots i_n\,k}(\omega)\sim \frac{1}{r}\left[1+\frac{\la}{r}+\dots \left(\frac{\la}{r}\right)^{n+1}\right]\left(\frac{d}{\la}\right)^{n+1}.
\eeqa
From equation \eref{ordF}) we can conclude that for the convergence of the expansion series it is necessary to have the inequality 
\beqa\label{ineq}
\la\,>\,d
\eeqa
satisfied. Consequently, for the spectral decomposition \eref{spectral}), the extension must be limited to the wave lengths verifying the last inequality.
We can conclude that there are three distinct regions around an isolate source with corresponding approximations of fields: the near (static) region  $d\,<r\,<<\,\la$, the intermediate (induction) region $r\approx \la$ and the far (radiation)  region $r>>\la$, Ref.~\cite{Jackson}. We consider here only  the last region.

\section{The radiation field } 
\label{wave region}
For $r\,>>\,\la$, the dominant term in the bracket from equation \eref{ordF}) is represented by the unity. The corresponding fields $\Evec$ and $\Bvec$ are proportional to $1/r$ and represent the parts contributing, as it is well known, to the energy and linear momentum radiated at large distances. The next term from the same bracket corresponds to the part of the field proportional to $1/r^2$ which contributes to the radiated angular momentum.  In the present paper, as in most textbooks of electrodynamics, we consider only the radiated energy and therefore, we calculate the part of the field proportional to $1/r$.  However, we point out that, as basic variables, the fields $\Evec$ and $\Bvec$ must be defined such that all observables of the radiated field can be calculated. The angular momentum is one of these observables (for a complete calculation of the radiation field, see Refs. \cite{RV}, \cite{cvcs} and literature cited in).

\par Let us calculate the first dominant terms, proportional to $1/r$, resulting from equations \eref{B1-0}) and \eref{E1-0}).  For this purpose, we retain  only the terms corresponding to the coefficients 
\beqa\label{Cn1}
 C^{(n,\,0)}_{i_1\dots i_n}=(-1)^n\,\nu_{i_1}\dots \nu_{i_n}.
 \eeqa
The expression from equation \eref{B1-0}) becomes 
\beqa\label{dB1a}
\Bvec_{\scriptstyle rad}(\rvec,t)=-\frac{\mu_0}{4\pi\al}\,\frac{1}{r}\,\nuvec\times\suml_{n\ge 0}\frac{1}{n!c^{n+1}}\,\frac{\d^{n+1}}{\d t^{n+1}}\left(\nuvec^n\vert\vert\mfrt^{(n+1)}(\tau_0)\right)
\eeqa
or 
\beqa\label{dB1a'}
\Bvec_{\scriptstyle rad}(\rvec,t)=-\frac{\mu_0}{4\pi\al}\,\frac{1}{r}\suml_{n\ge 0}\frac{1}{n!c^{n+1}}\,\left(\nuvec\times\frac{\d^{n+1}}{\d t^{n+1}}\mifrt(\tau_0;\nuvec,n+1)\right),
\eeqa
where we introduced the vector 
\beqa\label{mivec}
\mifrt(t;\xivec,n)=\xivec^{n-1}\vert\vert\mfrt^{(n)}(t).
\eeqa
Similarly, equation \eref{E1-0}) gets  
\beqa\label{dE1a}
\Evec_{\scriptstyle rad}(\rvec,t)
&=&\frac{1}{4\pi\eps_0}\,\frac{1}{r}\suml_{n\ge 1}\frac{1}{n! c^{n+1}}\frac{\d^{n+1}}{\d t^{n+1}}\left(\nuvec^n\vert\vert\psft^{(n)}(\tau_0)\right)\,\nuvec
\nonumber\\
&&-\frac{\mu_0}{4\pi\al^2}\,\frac{1}{r}\suml_{n\ge0}\frac{1}{n!c^n}\,\frac{\d^{n+1}}{\d t^{n+1}}\left(\nuvec^n\vert\vert\mfrt^{(n+1)}(\tau_0)\right),
\eeqa
since the source is an electric neutral system ($\dot{\psft}^{(0)}=0$), or\,  
\beqa\label{dE1a'}
\Evec_{\scriptstyle rad}(\rvec,t)&=&\frac{1}{4\pi\eps_0}\suml_{n\ge 1}\frac{1}{n!c^{n+1}}\frac{\d^{n+1}}{\d t^{n+1}}\left(\nuvec\cdot\bpi(\tau_0;\nuvec,n)\right)\,\nuvec\nonumber\\
&-&\frac{\mu_0}{4\pi\al^2}\,\frac{1}{r}\suml_{n\ge0}\frac{1}{n!c^n}\,\frac{\d^{n+1}}{\d t^{n+1}}\mifrt(\tau_0;\nuvec,n+1),
\eeqa
with  
\beqa\label{pivec}
\bpi(t;\xivec,n)=\xivec^{n-1}\vert\vert\psft^{(n)}(t).
\eeqa
Based on equation \eref{1}), we can write the following relation for the vector $\mifrt$: 
\beqa\label{1-m}
\mifrt(\tau_0;\nuvec,n+1)=-\al\,\nuvec\times\bmu(\tau_0;\nuvec,n)+\frac{1}{n+1}\dot{\bpi}(\tau_0;\nuvec,n+1),
\eeqa
where 
\beqa\label{muvec}
\bmu(t;\xivec,n)=\xivec^{n-1}\vert\vert\msft^{(n)}(t).
\eeqa
Substituting equation \eref{1-m}) in equation \eref{dB1a'}), one gets 
\beqa\label{dB1b}
\Bvec_{\scriptstyle rad}(\rvec,t)&=&\frac{\mu_0}{4\pi}\,\frac{1}{r}\left\{\suml_{n\ge 1}\frac{1}{n!c^{n+1}}\frac{\d^{n+1}}{\d t^{n+1}}\left[\nuvec\times\big(\nuvec\times\bmu(\tau_0;\nuvec,n)\big)
\right]\right.\nonumber\\
&& - \left.\frac{1}{\al}\suml_{n\ge 0}\frac{1}{(n+1)!c^{n+1}} \frac{\d^{n+2}}{\d t^{n+2}} \left[\nuvec\times\bpi(\tau_0;\nuvec,n+1)\right]    \right\}.
\eeqa
The change $n\rightarrow n+1$ in the summation index of the second sum allows us to write:
 \beqa\label{dB1a''}
\Bvec_{\scriptstyle rad}(\rvec,t)=\frac{\mu_0}{4\pi}\,\frac{1}{r}\nuvec\times\suml_{n\ge 1}\frac{1}{n!\,c^{n+1}}\frac{\d^{n+1}}{\d t^{n+1}}\left[\nuvec\times\bmu(\tau_0;\nuvec,n)-\frac{c}{\al}\bpi(\tau_0;\nuvec,n)\right].
\eeqa
Applying the same procedure to equation \eref{dE1a'}), we obtain  
\beqa\label{dE1b}
\Evec_{\scriptstyle rad}(\rvec,t)&=&\frac{1}{4\pi\eps_0}\,\frac{1}{r}\left\{\suml_{n\ge 1}\frac{1}{n!c^{n+1}}\frac{\d^{n+1}}{\d t^{n+1}}\left[\big(\nuvec\cdot\bpi(\tau_0;\nuvec,n)\big)\,\nuvec+\frac{\al}{c}\nuvec\times\bmu(\tau_0;\nuvec,n)\right]
\right.\nonumber\\
&&-\left.\frac{1}{c}\suml_{n\ge `0}\frac{1}{(n+1)!c^{n+1}}\frac{\d^{n+2}}{\d t^{n+2}}\bpi(\tau_0;\nuvec,n+1)\right\}.
\eeqa
The same change of the summation index $n\rightarrow n+1$ in the last sum gives:
\beqa\label{dE1a''}
\Evec_{\scriptstyle rad}(\rvec,t)=\frac{1}{4\pi\eps_0}\,\frac{1}{r}\nuvec\times\suml_{n\ge 1}\frac{1}{n!c^{n+1}}\frac{\d^{n+1}}{\d t^{n+1}}\left[\nuvec\times\bpi(\tau_0;\nuvec,n)+\frac{\al}{c}\bmu(\tau_0;\nuvec,n)\right].
\eeqa
Comparing equations \eref{dB1a''}) and  \eref{dE1a''}), we notice the  relation:
 \beqan
 \Evec_{\scriptstyle rad}(\rvec,t)\stackrel{\bpi(\tau_0;\nuvec,n)\leftrightarrow \bmu(\tau_0;\nuvec,n)}{ \longleftrightarrow}\Bvec_{\scriptstyle rad}(\rvec,t).
 \eeqan
As seen from equations \eref{dB1a''}) and \eref{dE1a''}), the fields $\Evec_{\scriptstyle rad}$ and $\Bvec_{\scriptstyle rad}$ are purely transverse fields satisfying the properties: 
\beqa\label{transv}
\nuvec\cdot\Evec_{\scriptstyle rad}=0,\;\;\nuvec\cdot\Bvec_{\scriptstyle rad}=0,\;\;
\Evec_{\scriptstyle rad}=\frac{c}{\al}\,\Bvec_{\scriptstyle rad}\times \nuvec,\;\;\eps_0\vert\Evec_{\scriptstyle rad}\vert^2=\frac{1}{\mu_0}\vert\Bvec_{\scriptstyle rad}\vert^2,
\eeqa
i.e. the characteristic wave properties.
\par Considering  the wave region, the essential parameter characterizing the multipole expansion is $d/\la$. Coming back to the monochromatic case studied in Section \ref{criterii}, and introducing the notation 
$\zeta=\frac{d}{\la}\,<<\,1$,
we can conclude that 
\beqa\label{59}
\frac{\d^k}{\d t^k}\,\psft^{(n)}\,\sim\,\left\{\begin{array}{c}\zeta^n,\;\;k\ge n\\\zeta^k,\;\;k\le n\end{array}\right.,\;\;\;\;\;
\frac{\d^k}{\d t^k}\,\msft^{(n)}\,\sim\,\left\{\begin{array}{c}\zeta^{n+1},\;\;k\ge n\\\zeta^{k+1},\;\;k\le n\end{array}\right. \ .
\eeqa
In the field multipolar expansions in terms of primitive moments $\psft^{(n)}$ and $\msft^{(n)}$ as, for example, in equations \eref{dB1b}) and \eref{dE1b}), the series terms appear in an increasing order in the parameter $\zeta$.
 If the multipolar series are expressed in terms of the {\bf STF}-tensors $\msftr^{(n)}$ and $\psftr^{(n)}$ as, for example, in equations \eref{dB1b}) and \eref{dE1b}), some caution is necessary. Replacing the pair $\msft^{(N-1)},\;\psft^{(N)}$ by the corresponding  {\bf STF}-projections, one induces in the moments of the inferior order compensating terms of orders $\zeta^k$, with $k\le N$. If we are interested in an approximation which includes the multipole moments up to a given rank $n\,<\,N$, we must discard the compensating terms from $\widetilde{\psft}^{(m)},\,\widetilde{\msft}^{(m)},\,m\le n$ of orders $\zeta^k$ for $k\,>\,n$.
 \par The task of writing finite multipolar sums in a well defined approximation becomes delicate when one considers products of field quantities as, for example, when defining the radiation intensities. The simplest, and most important, is the case of the radiated power. The Poynting vector can be written in terms of $\Evec_{\scriptstyle rad}$ and $\Bvec{\scriptstyle rad}$ with the help of equations \eref{transv}) as: 
 \beqa\label{Poynting}
 \bsy{S}=\frac{\al}{\mu_0}(\Evec\times\Bvec)=\eps_0\vert\Evec_{\scriptstyle rad}\vert^2\,c\nuvec+{\cal O}(\frac{1}{r^3})
 =\frac{1}{\mu_0}\vert\Bvec_{\scriptstyle rad}\vert^2\,c\nuvec+{\cal O}(\frac{1}{r^3}).
 \eeqa
 $\Bvec$ and $\Evec$ are considered real vectors.
 The total radiated power may be written as the limit of a surface integral on a sphere centered in $O$, of radius $r$, for $r\to \infty$. Let us express the energy current in the radiation approximation corresponding to the sphere of radius $r$: 
 \beqa\label{NW} N(\bsy{S},\Sigma_r;t)=\oint_{\Sigma_r}\,r^2\nuvec\cdot\bsy{S}(\rvec,t)\,\rmd\Omega(\nuvec)=\frac{c}{\mu_0}\oint_{\Sigma_r}\,r^2\vert\Bvec(\rvec,t)\vert^2\,\rmd\Omega(\nuvec)\,+\,{\mathcal O}(\frac{1}{r}).
 \eeqa
 The integrand from the last expression represents the angular distribution of the radiation. The quantity $N(\bsy{S},\Sigma_r;t)\,\rmd t$  represents the energy which crosses the sphere $\Sigma_r$ in the time interval $(t,\,t+\rmd t)$ and is determined by the values of the multipole moments of the source in the interval $(t-r/c,\,t+\rmd t-r/c)$. For large, but finite $r$, $r>>\la$, this is an approximate expression obtained by neglecting the terms of order at least $1/r$. The expression from equation \eref{NW}) can be employed for drawing conclusions on the electric charge distribution at the retarded time from observations, on the angular distribution or on the total radiated power. Since  $\rmd t=\rmd\tau_0$, we can say that $N(\bsy{S},\Sigma_r;t)\,\rmd t$ is the part of the energy emitted by the source in the given time interval which contributes to the energy intensity corresponding to $\Sigma_r$ at the time $t+r/c$. If in equation \eref{NW}) we put $t$ instead $\tau_0$, then the quantity $\lim_{r\to \infty} N(\bsy{S},\Sigma_r;t)\, \rmd t$
 represents that part of the energy emitted by the source which contributes to the radiated energy or, shortly, radiated by the source. The situation changes when the support of the source depends on time, in particular, for the radiation of a moving poit-like source (see Ref. \cite{Landau},$\&$73).
 \par For calculating the angular distribution of the energy radiated or the total radiated power, one deals with the square of $\Bvec_{\scriptstyle rad}$ or  $\Evec_{\scriptstyle rad}$. Let us consider firstly that we operate with the series \eref{dB1b}) or \eref{dE1b}) expressed in terms of primitive electric and magnetic moments . For illustrating the kind of problems which appear in the approximation process, it is sufficient to consider the well known problem of the calculation of the electric and magnetic dipole and electric quadrupole field contributions to the radiation (see Refs. \cite{Landau}, \cite{Jackson}). From the very beginning, we point out that here, this calculation is associated with the problem of the radiation of a complex system characterized by an arbitrary number of electric and magnetic multipoles. Once considered the contribution of the electric quadrupole field, from the expression of $\vert\Bvec_{\scriptstyle rad}\vert^2$, with  $\Bvec$ considered real, the square of the electric quadrupolar moment $\psft^{(2)}$ is  part of this contribution. This is a term of the order $\zeta^4$. There is a similar conclusion concerning the contribution of the magnetic dipolar moment $\mvec$. It is easy to see that in a consistent approximation procedure, together with  the squares of  $\psft^{(2)}$,  $\mvec$  and $\pvec$, we must consider also the products of the electric dipolar moment $\pvec$ with the electric octopolar moment $\psft^{(3)}$ and the magnetic quadrupolar moment $\msft^{(2)}$. They also give contributions of the order $\zeta^4$. Let us estimate these contributions in terms of primitive moments. In equation \eref{Poynting}), to the contribution of $(\Bvec^{(0)}_{\scriptstyle rad}+\Bvec^{(1)}_{\scriptstyle rad})^2$, we have to add the fourth-order contribution of the product $\Bvec^{(0)}_{\scriptstyle rad}\cdot\Bvec^{(2)}_{\scriptstyle rad}$.
From equation \eref{dB1b})  we can write, for $n=0,\,1,\,2$: 
\beqa\label{B(0)}
\Bvec^{(0)}_{\scriptstyle rad}(\rvec,t)=-\frac{\mu_0}{4\pi\al c}\,\frac{1}{r}\,\nuvec\times\ddot{\bpi}(\tau_0;\nuvec,1),
%=-\frac{\mu_0}{4\pi\al c}\,\frac{\nuvec\times\ddot{ \pvec}(\tau_0)}{r},
\eeqa
\beqa\label{B(1)}
\Bvec^{(1)}_{\scriptstyle rad}(\rvec,t)&=&\frac{\mu_0}{4\pi c^2}\,\frac{1}{r}\left\{\nuvec\times\big[\nuvec\times\ddot{\bmu}(\tau_0;\nuvec,1)\big]-\frac{1}{2\al} \,\nuvec\times\tdot{\bpi}(\tau_0;\nuvec,2)\right\},
\eeqa
\beqa\label{B(2)}
\Bvec^{(2)}_{\scriptstyle rad}(\rvec,t)=\frac{\mu_0}{4\pi c^3}\,\frac{1}{2r}\,\left\{\nuvec\times\big[\nuvec\times\tdot{\bmu}(\tau_0;\nuvec,2)\big]-\frac{1}{3\al}\nuvec\times\qdot{\bpi}(\tau_0;\nuvec,3)\right\}.
\eeqa
The angular distribution and the total power of the radiation are given by 
\beqa\label{I(niu)}
{\mathcal I}(\nuvec)=\frac{c}{\mu_0}r^2\,\big(\Bvec_{\scriptstyle rad}\big)^2,\;\;\;
{\mathcal I}=\frac{4\pi c}{\mu_0}\langle r^2\big(\Bvec_{\scriptstyle rad}\big)^2\rangle,
\eeqa
where 
\beqa
\langle\,f(\nuvec)\,\rangle=\frac{1}{4\pi}\int f(\nuvec)\,\rmd\Omega(\nuvec).
\eeqa
Considering, for example, the angular distribution ${\mathcal I}(\nuvec)$, expanded up to the order $\zeta^4$, we have  
\beqa\label{Iniu4}
{\mathcal I}(\nuvec)=\frac{c}{\mu_0}r^2\big[\big(\Bvec^{(0)}{\scriptstyle rad} + \Bvec^{(1)}_{\scriptstyle rad}\big)^2 + 2\Bvec^{(0)}_{\scriptstyle rad}\cdot\Bvec^{(2)}_{\scriptstyle rad}\big] +{\mathcal O}(\zeta^5).
\eeqa
In the following, for simplifying the notation, we understand by $\bpi(n)$ and $\bmu(n)$ the corresponding vectors with the arguments $\tau_0$ and $\nuvec$. 
A simple algebraic calculation gets: 
\beqa\label{B0B1}
&&\frac{c}{\mu_0}\,r^2\big(\Bvec^{(0)}_{\scriptstyle rad}+\Bvec^{(1)}_{\scriptstyle rad}\big)^2=\frac{1}{(4\pi)^2\eps_0c^3}\left[\big(\nuvec\times\ddot{\bpi}(1)\big)^2+\frac{\al^2}{c^2}\big(\nuvec\times\ddot{\bmu}(1)\big)^2+\frac{1}{4c^2}\big(\nuvec\times\tdot{\bpi}(2)\big)^2\right.\nonumber\\
&&\hspace{0.5cm}+\left.\frac{2\al}{c^2}\ddot{\bmu}(1)\cdot\big(\nuvec\times\tdot{\bpi}(2)\big)+\frac{1}{c}\big(\ddot{\bpi}(1)\cdot\tdot{\bpi}(2)\big)-\frac{1}{c}\big(\nuvec\cdot\ddot{\bpi}(1)\big)\big(\nuvec\cdot\tdot{\bpi}(2)\big)
\right],
\eeqa
\beqa\label{B0B2}
\frac{2c}{\mu_0}\,r^2\,\Bvec^{(0)}_{\scriptstyle rad}\cdot \Bvec^{(2)}_{\scriptstyle rad}&=&\frac{1}{(4\pi)^2\eps_0 c^5}\left[\al \tdot{\bmu}(2)\cdot\big(\nuvec\times\ddot{\bpi}(1)\big)+\frac{1}{3}\big(\ddot{\bpi}(1)\cdot\qdot{\bpi}(3)\big)
\right.\nonumber\\
&&-\left.\frac{1}{3}\big(\nuvec\cdot\ddot{\bpi}(1)\big)\big(\nuvec\cdot\qdot{\bpi}(3)\big)\right].
\eeqa
One can determine the averaged quantities from the last two equations using  formula \cite{Thorne}: 
\beqa\label{numed}
\langle\nu_{i_1}\dots\nu_{i_n}\rangle=\left\{\begin{array}{c}0,\;\;\;\;\;\;\;\;\;\;\;\;\;\;\;\;\;\;\;\;\;\;\;\;\;\;n=2k+1,\\
\frac{1}{(2k+1)!!}\,\delta_{\{i_1i_2}\dots\delta_{i_{n-1}i_n\}},\;\;\;\;\;\;\;\;\;\;n=2k,\;\;\;\;\;k=0,1,\dots\end{array}\right. .
\eeqa
The averaged expressions from equations \eref{B0B1}) and \eref{B0B2}) are given by  
\beqa\label{B0B1'}
\frac{c}{\mu_0}\,r^2\big{\langle}\big(\Bvec^{(0)}_{\scriptstyle rad}+\Bvec^{(1)}_{\scriptstyle rad}\big)^2\big{\rangle}=
\frac{1}{4\pi\eps_0c^3}\left(\frac{2}{3}\,\ddot{\pvec}^2+\frac{2\al^2}{3c^2}\,\ddot{\mvec}^2+\frac{1}{20c^2}\tdot{\psf}_{ij}\,\tdot{\psf}_{ij}-\frac{1}{60\,c^2}\tdot{\psf}^2_{ii}
\right)
\eeqa
and  
\beqa\label{B0B2'}
\frac{2c}{\mu_0}\,r^2\,\big{\langle}\Bvec^{(0)}_{\scriptstyle rad}\cdot \Bvec^{(2)}_{\scriptstyle rad}\big{\rangle}=
\frac{1}{4\pi\eps_0c^5}\left(-\frac{\al}{3}\,\ddot{p_k}\,\eps_{kij}\tdot{\msf}_{ij}+ \frac{2}{45}\,\ddot{p}_k\qdot{\psf}_{kjj}
\right).
\eeqa
In equations \eref{B0B1'}) and \eref{B0B2'}) some standard notation  for magnetic and electric moments is used. A simple calculation shows that  
\beqa\label{It}
\frac{2c}{\mu_0}\,r^2\,\big{\langle}\Bvec^{(0)}_{\scriptstyle rad}\cdot \Bvec^{(2)}_{\scriptstyle rad}\big{\rangle}=
-\frac{1}{4\pi\eps_0\,c^3}\,\frac{4}{3c^2}\,\ddot{\pvec}\,\cdot\!\tdot{\bsy{t}},
\eeqa
where $\bsy{t}$ is defined by equation \eref{deft}). The total radiated power is then given by 
\beqa\label{tpower}
{\mathcal I}=\frac{1}{4\pi\eps_0c^3}\left(\frac{2}{3}\,\ddot{\pvec}^2+\frac{2\al^2}{3c^2}\,\ddot{\mvec}^2+\frac{1}{20c^2}\tdot{\psf}_{ij}\,\tdot{\psf}_{ij}-\frac{1}{60\,c^2}\tdot{\psf}^2_{ii}-\frac{4}{3c^2}\,\ddot{\pvec}\,\cdot\!\tdot{\bsy{t}}\right).
\eeqa
Considering in equation \eref{tpower}) the multipolar moments at the time $t$, we have the description of the source emission. This expression differs from the result given in Ref. \cite {Landau}  firstly by the fact that, whereas here the electric 4-polar moment is the primitive one, in this reference it is represented by the corresponding {\bf STF} tensor and, consequently, the term containing the trace of the electric 4-polar moment is absent.
 The second difference consists in the presence in equation \eref{tpower}) of the  toroidal moment $\bsy{t}$. 
 \par For expressing ${\mathcal I}$ by {\bf STF}- tensors, we apply the invariance of the field to the substitutions of all multipole primitive tensors $\psft$ and $\msft$ by  corresponding {\bf STF}-tensors $\widetilde{\psft}$, $\widetilde{\msft}$ . These last tensors are, in general, different from the correspondent {\bf STF} projections as it was shown in Section \ref{STF}. Obviously, for the approximation corresponding to equation \eref{tpower}), it is sufficient to consider the {\bf STF}-projections $\widetilde{\psft}^{(3)}=\Tcal\big(\psft^{(3)}\big)$ and $\widetilde{\msft}^{(2)}=\Tcal\big(\msft^{(2)}\big)$ and the induced transformations $\psft^{(k)}\longrightarrow \widetilde{\psft}^{(k)}$ for $k=1,2$,  and   $\msft^{(k)}\longrightarrow \widetilde{\msft}^{(k)}$ for $k=1$ (see Section \ref{STF}, equations \eref{trp3})-\eref{deft})). Performing these substitutions in equations \eref{B0B1'}) and \eref{B0B2'}), we obtain for ${\mathcal I}$:
 \beqa\label{tpower'}
 {\mathcal I}=\frac{1}{4\pi\eps_0c^3}\left(\frac{2}{3}\,(\ddot{\pvec}-\frac{1}{c^2}\tdot{\bsy{t}})^2+\frac{2\al^2}{3c^2}\,\ddot{\mvec}^2+\frac{1}{20c^2}\tdot{\pisf}_{ij}\,\tdot{\pisf}_{ij}\right) + {\mathcal O}(\zeta^5).
\eeqa
The term $\bsy{t}^2/c^4$ must be eliminated since it is of order $\zeta^6$, such that the final expression of ${\mathcal I}$ in the required approximation is given by  
\beqa\label{tpower''}
{\mathcal I}=\frac{1}{4\pi\eps_0c^3}\left(\frac{2}{3}\,\ddot{\pvec}^2+\frac{2\al^2}{3c^2}\,\ddot{\mvec}^2+\frac{1}{20c^2}\tdot{\pisf}_{ij}\,\tdot{\pisf}_{ij}-\frac{4}{3c^2}\,\ddot{\pvec}\,\cdot\!\tdot{\bsy{t}}\right).
\eeqa
This is the correct expression of the radiated power in the $\zeta^4$-approximation and it is obviously not represented by independent contributions of $\pvec,\,\mvec, \pisft^{(2)}$.
 In Ref. \cite{Jackson}, the contribution of the  multipole moment  $\psft^{(2)}$ is calculated as an independent contribution. The corresponding result is a correct one if the radiation source is an elementary system characterized only by his electric 4-polar moment. The reader must be warned that in case one considers a complex system characterized by higher multipole moments, the correct evaluation of the contributions of the moments $\pvec,\,\mvec$ and $\psft^{(2)}$ cannot be identified with the sum of the independent contributions of these multipole moments. Particularly, this is the reason why in Ref. \cite{Landau} (\& 71) one obtains an incomplete result when calculating  the electric and magnetic dipolar and electric quadripolar radiation considered as a result of the multipole expansion of a complex system.

\par A simpler procedure for obtaining the expression from equation \eref{tpower''}) is achieved if in equations \eref{B0B1}) and \eref{B0B2}) one  substitutes $\bpi(t;\nuvec,n)\rightarrow \tilde{\bpi}(t;\nuvec,n)=\nuvec^{n-1}\vert\vert\widetilde{\psft}^{(n)}(t)$ and $\bmu(t;\nuvec,n)\rightarrow \tilde{\bmu}(t;\nuvec,n)=\nuvec^{n-1}\vert\vert\widetilde{\msft}^{(n)}(t)$. This procedure has the advantage of a  simpler calculation of the $\nuvec$-averaged quantities for higher order terms of ${\mathcal I}$.  In appendix \ref{sec:formulae} we give some general results for the radiated energy and linear and angular momenta (for detailed calculations see Ref. \cite{cvcs} and the issues cited there).

\section{Translational invariance}\label{translation} 
Let us consider a new  point of reference $O'$ specified in the Cartesian reference system $\{O,\,(\evec_i)_{i=1\div 3}\}$ by the vector $\avec=a_i\,\evec_i$. The new reference system in the affine space is defined as $\{O',\,(\evec_i)_{i\div 3}\}$.  Here, both  systems  are associated with the same basis in the vectorial space. For avoiding a possible confusion produced by the notation used, we denote by $P$ the point where  the field is calculated and by $P'$ the current integration point. Correspondingly, instead of  $\Bvec(\rvec,t)$ and $J_k(\rvec',\tau)$ in equation \eref{B1}), the notation  $\Bvec(P,t)$ and $J_k(P',\tau)$ will be used. Generally, the multipole moments depend on the point of reference chosen. As it is stressed in Ref. \cite{Visschere}, the ``origin dependence'' of multipole moments forms only a part of the effect of changing the point of reference. Considering consistently all the  the changes  in the calculation of an   observable of the physical system,  one can verify the invariance of this observable. 
We can write equation \eref{B1}) in the new system of reference as 
\beqa\label{B1-O1}
\Bvec'(P,t)
=\frac{\mu_0}{4\pi\al}\evec_i\eps_{ijk}\intl_{\dom}\left(\d_j\frac{J_k(P',t-\frac{\vert\rvec-\avec-\xivec\vert}{c})}{\vert\rvec-\avec-\xivec\vert}\right)_{\xivec=\rvec''}\rmd^3x'',
\eeqa
where $\rvec''=x''_i\,\evec_i$ is the vector associated with the pair  $O'P'$. Writing the Taylor series about $\xivec=0$ and, after that, performing the substitution $\xivec=\rvec''$, one gets
\beqan
\!\!\!\!\!\!\!\!\!\!\!\!\Bvec'(P,t)&=&\frac{\mu_0}{4\pi\al}\,\evec_i\eps_{ijk}\suml_{n\ge 0}\frac{(-1)^n}{n!}
\intl_{\dom}\,x''_{i_1}\dots x''_{i_n}\d_j\d_{i_1}\dots\d_{i_n}\frac{J_k(P',t-\frac{\vert\rvec-\avec\vert}{c})}{\vert\rvec-\avec\vert}\rmd^3x''\\
\!\!\!\!\!\!\!\!\!\!\!\!&=&\frac{\mu_0}{4\pi\al}\,\evec_i\eps_{ijk}\suml_{n\ge 0}\frac{(-1)^n}{n!}\intl_{\dom}(x'_{i_1}-a_{i_1})\dots (x'_{i_n}-a_{i_n})
\d_j\,\d_{i_1}\dots \d_{i_n}\frac{J_k(P',t-\frac{\vert\rvec-\avec\vert}{c})}{\vert\rvec-\avec\vert}\rmd^3x'\\
\eeqan
or
\beqan
\!\!\!\!\!\!\!\!\!\!\!\!\Bvec'(P,t)&=&\frac{\mu_0}{4\pi\al}\,\evec_i\eps_{ijk}\suml_{n\ge 0}\frac{(-1)^n}{n!}\suml^n_{m=0}(-1)^m\,\frac{n!}{m!(n-m)!}\,a_{i_1}\dots a_{i_m}\\
\!\!\!\!\!\!\!\!\!\!\!\!&\times& \intl_{\dom}\d_j\,\d_{i_1}\dots \d_{i_n}\,x'_{i_{m+1}}\dots x'_{i_n}\frac{J_k(P',t-\frac{\vert\rvec-\avec\vert}{c})}{\vert\rvec-\avec\vert}\rmd^3x'.
\eeqan
Performing the Taylor expansion of the fraction $J_k(P',t-\vert\rvec-\avec\vert/c)/\vert\rvec-\avec\vert$ as a function of $\avec$ about $\avec=0$, it results: 
\beqa\label{B1-O2}
\!\!\!\!\!\!\!\Bvec'(\rvec,t)&=&\frac{\mu_0}{4\pi\al}\,\evec_i\eps_{ijk}\suml_{n\ge 0}(-1)^n\suml^n_{m=0}\frac{(-1)^m}{m!(n-m)!} a_{i_1}\dots a_{i_m}\suml_{l\ge 0}\frac{(-1)^l}{l!}a_{j_1}\dots a_{j_l}\nonumber\\
 \!\!\!\!\!\!\!&&\times \d_j\,\d_{i_1}\dots \d_{i_n}\,\d_{j_1}\dots\d_{j_l}
\left(\frac{1}{r}\intl_{\dom}x'_{m+1}\dots x'_{i_n}J_k(\rvec',t-\frac{r}{c})\right)\,\rmd^3x'.
\eeqa
This equation can be written, in a more compact form, as:  
\beqa\label{B1-O3}
\!\!\!\!\!\!\!\!\!\!\!&&\Bvec'(\rvec,t)\nonumber\\
&&=\frac{\mu_0}{4\pi\al}\,\evec_i\eps_{ijk}\d_j\suml_{n\ge 0}\suml^n_{m=0}\frac{(-1)^{n-m}}{m!(n-m)!}(\avec\cdot\nablav)^m
\suml_{l\ge 0}\frac{(-1)^l}{l!}(\avec\cdot\nablav)^l\,\left(\nablav^{n-m}\vert\vert\frac{\mfrt^{(n-m+1)}(\tau_0)}{r}\right)_k\nonumber\\
\!\!\!\!\!\!\!\!\!\!\!&&=\frac{\mu_0}{4\pi\al}\,\evec_i\eps_{ijk}\d_j\suml_{n\ge 0}\suml^n_{m=0}\frac{(-1)^{n-m}}{m!(n-m)!}(\avec\cdot\nablav)^m
\rme^{-\avec\cdot\nablav}\left(\nablav^{n-m}\vert\vert\frac{\mfrt^{(n-m+1)}(\tau_0)}{r}\right)_k.
\eeqa
Obviously, considering the infinite series corresponding to the multipole expansion of the magnetic field, we have no dependence of the choice of the reference point since the Taylor series gives the same values of a function no matter what is the point about the expansion is taken. Indeed, considering  the last expression of $\Bvec'$, for the sums after  $n$ and $m$, we have:
\beqan
\Bvec'(\rvec,t)&=&\suml^\infty_{n=0}\suml^n_{m=0}\bsy{f}_{nm}=\suml^\infty_{n=0}\suml^\infty_{m=0}\theta(n-m)\bsy{f}_{nm}
=\suml^\infty_{m=0}\suml^\infty_{n=m}\bsy{f}_{nm},
\eeqan  
i.e.\ the possibility of inverting the order of the summation after $n$ and $m$ in the case of infinite series. Using this property in equation \eref{B1-O3}) and introducing a new summation index $q=n-m$, we obtain
\beqan
\!\!\!\!\!\!\!\!\!\!\!&&\Bvec'(\rvec,t)=\frac{\mu_0}{4\pi\al}\,\evec_i\eps_{ijk}\d_j\,\suml_{m\ge 0}\frac{(\avec\cdot\nablav)^m}{m!}
\suml_{q\ge0}\frac{(-1)^q}{q!}\rme^{-\avec\cdot\nablav}\left(\nablav^q\vert\vert\frac{\mfrt^{(q+1)}(\tau_0)}{r}\right)_k.
\eeqan
Recognizing in the last expression the series of  $exp(\avec\cdot\nablav)$ and the expansion \eref{B1-2}) of $\Bvec(\rvec,t)$, one can conclude that $\Bvec'=\Bvec$, i.e.\ the translation invariance of the magnetic field multipole expansion. For practical calculations, we have to take into account  the expression in equation \eref{B1-O3}) as a double series  expansion upon the parameters $\avec$ and $d$.  Since $a\,\leq\,d$, the parameters $\avec$ and $d$ are associated with the subunit and adimensional parameter $\zeta=d/\la\sim a/\la$. Obviously, the translation invariance cannot be verified by each term of the field expansion but, for a given approximation defined by a maximum power of the parameter $\zeta$, it is verified for a well defined partial sum from the series. For illustrating how this invariance property is working, we write equation \eref{B1-O3}) as
\beqan
\Bvec'(\rvec,t)
=\suml_{n\ge 0}\suml^n_{m=0}\frac{(\avec\cdot\nablav)^m}{m!}\,\rme^{-\avec\cdot\nablav}\Bvec^{(n-m)}(\rvec,t).
\eeqan
Let us consider the first three terms from the field series:
\beqan
\!\!\!\!\!\!\!\!\Bvec^{'(0)}&=&\,\rme^{-\avec\cdot\nablav}\Bvec^{(0)}
=\big[1-(\avec\cdot\nablav)+\frac{1}{2}(\avec\cdot\nablav)^2-\frac{1}{6}(\avec\cdot\nablav)^3+\dots\big]\Bvec^{(0)},\\
\!\!\!\!\!\!\!\!\Bvec^{'(1)}&=&\big[(\avec\cdot\nablav)-(\avec\cdot\nablav)^2+\frac{1}{2}(\avec\cdot\nablav)^3+\dots\big]\,\Bvec^{(0)}
+\big[1-(\avec\cdot\nablav)+\frac{1}{2}(\avec\cdot\nablav)^2+\dots\big]\Bvec^{(1)},\\
\!\!\!\!\!\!\!\!\Bvec^{'(2)}&=&\big[\frac{1}{2}(\avec\cdot\nablav)^2-\frac{1}{2}(\avec\cdot\nablav)^3+\dots\big]\Bvec^{(0)}
+\big[(\avec\cdot\nablav)-(\avec\cdot\nablav)^2+\dots\big]\Bvec^{(1)}\nonumber\\
\!\!\!\!\!\!\!\!&&+\big[1-(\avec\cdot\nablav)+\dots\big]\Bvec^{(2)}.
\eeqan
Since $(\avec\cdot\nablav)$ is associated to the parameter $\zeta$ and $\Bvec^{(0)}$, $\Bvec^{(1)}$ and $\Bvec^{(2)}$ 
are of the orders $\zeta,\,\zeta^2$ and $\zeta^3$, respectively, we can express the partial sums up to the order $\zeta^4$ inclusively: 
\beqa\label{B0B11}
\Bvec^{'(0)}+\Bvec^{'(1)}&=&\Bvec^{(0)}+\Bvec^{(1)}
+\big[-\frac{1}{2}(\avec\cdot\nablav)^2+\frac{1}{3}(\avec\cdot\nablav)^3\big]\Bvec^{(0)}\nonumber\\
&+&\big[-(\avec\cdot\nablav)+\frac{1}{2}(\avec\cdot\nablav)^2\big]\Bvec^{(1)}+\mathcal{O}(\zeta^5), 
\eeqa
\beqa\label{B0B21}
\Bvec^{'(0)}+\Bvec^{'(1)}+\Bvec^{'(2)}&=&\Bvec^{(0)}+\Bvec^{(1)}+\Bvec^{(2)}\nonumber\\
&-&\frac{1}{6}(\avec\cdot\nablav)^3\Bvec^{(0)}-\frac{1}{2}(\avec\cdot\nablav)^2\Bvec^{(1)}-(\avec\cdot\nablav)\Bvec^{(2)}+\mathcal{O}(\zeta^5).
\eeqa
From equations \eref{B0B11})  and \eref{B0B21}) it is seen that the sum $\Bvec^{(0)}+\Bvec^{(1)}$ is translational invariant up to the order $\zeta^2$ and the sum $\Bvec^{(0)}+\Bvec^{(1)}+\Bvec^{(2)}$, up to the order $\zeta^3$. An interesting circumstance appears in case one analyzes the second powers of these sums as, for example, when one calculates the radiation intensities. Let us consider the squared sum from equation \eref{B0B11}): 
\beqa\label{B0B3}
\big(\Bvec^{'(0)}+\Bvec^{'(1)}\big)^2=\big(\Bvec^{(0)}+\Bvec^{(1)}\big)^2+\mathcal{O}(\zeta^4).
\eeqa
Terms of the orders $\zeta^3$ and $\zeta^4$ are present in $\big(\Bvec^{(0)}+\Bvec^{(1)}\big)^2$ such that we have to ensure the translational invariance up the order four in the parameter $\zeta$. Therefore, the contribution of  $\Bvec^{'(2)}$ must be added up since this term introduces the product $\Bvec^{(0)}\cdot\Bvec^{(2)}\sim \zeta^4$: 
\beqa\label{B0B4}
\big(\Bvec^{'(0)}+\Bvec^{'(1)}+\Bvec^{'(2)}\big)^2=\big(\Bvec^{(0)}+\Bvec^{(1)}\big)^2
+2\Bvec^{(0)}\cdot\Bvec^{(2)}+\mathcal{O}(\zeta^5).
\eeqa
\par In terms of the multipole moments $\mfrt$, we conclude that, if one includes the square of the magnetic moment $\mfrt^{(1)}$ then, for ensuring the translation invariance, the inclusion of a partial  contribution of $\mfrt^{(2)}$ 
is necessary, too. This partial contribution is expressed in terms of the contraction of the tensors $\mfrt^{(2)}$ and $\mfrt^{(0)}$. Correspondingly, including the squares of the electric 4-polar moment $\psft^{(2)}$ and of the magnetic dipolar moment $\mvec$ we have to include also partial contributions of the moments $\msft^{(2)}$ and $\psft^{(3)}$ represented by contractions of the corresponding tensors with the electric dipolar moment $\pvec$. This property explains why in Ref. \cite{Zeldovich}, introducing in the expression of the radiated power only the sum of the moments $\pvec$, $\mvec$ and $\psft^{(2)}$ squared, finally, for establishing  a translation invariant result one performs a correction by a term represented by contractions of $\msft^{(2)}$ and $\psft^{(3)}$ with $\pvec$. In such a way one introduces in fact a contribution of the dipole toroidal moment $\bsy{t}$. 
\par The procedure can be continued for higher orders. In the following step, for example, the contribution of the square of $\Bvec^{(0)}+\Bvec^{(1)}+\Bvec^{(2)}$ up to the order $\zeta^6$ is obtained as 
\beqa\label{B0B5}
\big(\Bvec^{'(0)}+\Bvec^{'(1)}+\Bvec^{'(2)}\big)^2
=\big(\Bvec^{(0)}+\Bvec^{(1)}+\Bvec^{(2)}\big)^2+2\big(\Bvec^{(0)}\cdot\Bvec^{(3)}+\Bvec^{(1)}\cdot\Bvec^{(3)}\big).
\eeqa
Besides the contributions of the multipole moments $\pvec,\,\mvec,\,\psft^{(2)},\msft^{(2)},\,\psft^{(3)}$, we have to include also some partial contributions from $\psft^{(4)}$ and $\msft^{(3)}$ represented by the last two terms from equation \eref{B0B5}).
\par Combining the above observations with the results of the previous section, we can conclude here that all the invariance properties are ensured if a consistent approximation procedure is used.  
 For this reason, we suspect some inconsistencies  in considering the approximation criteria when,  performing multipole expansion, one  concludes that a correction is necessary for ensuring the translational invariance(see Refs. \cite{Zeldovich},\cite{Visschere}, \cite{Raab}).
\section{Conclusion}\label{conclusion}

Along this paper, we mainly exposed the fundamental grounds for the calculus procedure of the multipole expansion of the electromagnetic field. 
\par Firstly, the direct approach in the expansion of the quantities $\Evec$ and $\Bvec$ from Jefimenko's equations revealed that this method is neither more difficult, nor easier than the one based on the expansion of the potentials $\Avec$ and $\Phi$. An advantage of the current approach can be the elimination, at least partly, of the difficulties related to the inversion of different operations, mostly of derivation and integration. Another advantage is offered by the direct examination of the invariance of the fields $\Evec$ and $\Bvec$ to the transformations required by the manipulation of the electric and magnetic moment tensors (sections \ref{Field expansion} and \ref{STF}). In this case, one does not need to employ gauge invariance.
\par Secondly, in section \ref{criterii}, we specify the approximation criteria according to which one has to consider three approximation intervals for the field associated to a localized source, as mentioned in  Refs. \cite{Jackson} and~ \cite{Landau}.  Then, in the next two sections, for radiation calculus, we consistently apply the criterion $d/\la$.  The consequences are the ones mainly known from Ref. \cite{Zeldovich}.
\par We hope that the discussion from section \ref{translation} regarding the invariance of the physical results when changing the point of reference, although without an extension to the general multipole terms, is convincing enough. We concluded that the translational invariance is automatically ensured if one respects the approximation criteria when employing multipole expansions. 
\par  Since in the main part of the current work we wanted to avoid the generalization to superior orders of the multipole series, in the Appendix we summarized for the interested reader, the main known results in the literature.  
\par We believe that in an electrodynamics lecture, there should exist some remarks on the contributions of the toroidal moments, at least for the quadrupole electric moment. In this approximation, as one can see from sections \ref{STF} and \ref{wave region}, the highlight of the contributions is not involving a considerable computational effort.

\appendix
\section{Some basic formulae for generalizing the procedure of reducing the electric and magnetic momentum tensors}
\label{sec:formulae} 
In this section we list the required formulae for the generalized tensor reducing procedure. The equations are given without the related proofs since they can be found in the literature (see Ref.~\cite{cvcs}).
\par Let $\Ssft^{(n)}$ be a symmetric tensor of rank  $n$. Its {\bf STF}- projection results from the following formula: 
\beqa\label{TSn}
\Tcal\big(\Ssft^{(n)}\big)_{i_1\dots i_n}=\Ssft_{i_1\dots i_n}-\delta_{\{i_1i_2}\lasf(\Ssft^{(n)})_{i_3\dots i_n\}}.
\eeqa
The operator  $\lasft$ is defined on the account of a formula for the  {\bf STF}-projection  of a symmetric tensor given in  Ref.~\cite{Thorne} (the book \cite{Pirani} is cited as the origin of this formula): 
\beqa\label{Lambda}
\lasf\big(\Ssft^{(n)}\big)_{i_1\dots i_{n-2}}&=&\suml^{[n/2-1]}_{m=0}\frac{(-1)^m[2n-1-2(m+1)]!!}{(m+1)(2n-1)!!}\nonumber\\
&&\times\,\delta_{\{ i_1i_2}\dots\delta_{i_{2m-1}i_{2m}}
 \Ssf^{(n;m+1)}_{i_{2m+1}\dots i_{n-2}\}}.
\eeqa
The proof can be found in Ref.~\cite{Apple}. The notation  $\Ssf^{(n;p)}$  indicates $p$ pairs of contracted indices.  In the following, for simplifying the notation, all arguments of the operator $\lasft$ should be considered symmetric tensors, i.e.\ $\lasft(\Tsft^{(n)})=\lasft(\Scal(\Tsft^{(n)}))$ for any tensor $\Tsft^{(n)}$. The same applies to the operator $\Tcal$: $\Tcal(\Tsft^{(n)})=\Tcal(\Scal(\Tsft^{(n)}))$, $\Scal$ being the symmetrization operator.
\par In the symmetrization  process we have to calculate the symmetric projections of some tensors $\Lsft^{(n)}$ of the magnetic moment type: they are  symmetric in the first $n-1$ indices and the contraction of $i_n$ with any index $i_q,\;q=1\dots n-1$ gives a null result. For the symmetric projection of such a tensor, we introduce the formula: 
\beqa\label{simL}
\big(\Scal(\Lsft^{(n)})\big)_{i_1\dots i_n}=\Lsf_{i_1\dots i_n}
-\frac{1}{n}\suml^{n-1}_{\la=1}\eps_{i_{\la}i_nq}\Ncal^{(\la)}_{i_1\dots i_{n-1}q}(\Lsft^{(n)}),
\eeqa
where $\Ncal^{(\la)}_{i_1\dots i_{n-1}}$ is the component with the $i_{\la}$ index  suppressed. The operator $\Ncal$ defines a correspondence between $\Lsft^{(n)}$ and a tensor of rank $(n-1)$ of the same type from  the symmetry point of view. Particularly, 
\beqa\label{N(M)}
&~&\Ncal^{2k}\big(\msft^{(n)}\big)=\frac{(-1)^k n}{(n+1)\al}\intl_{\dom}(r^2)^k\,\rvec^{n-2k}\times\jvec\,\rmd^3x,\nonumber\\
&~&\Ncal^{2k+1}\big(\msft^{(n)}\big)=\frac{(-1)^kn}{(n+1)\al}
\intl_{\dom}(r^2)^k\,\rvec^{n-2k-1}\times(\rvec\times\jvec)\,\rmd^3x,\;\;k=0,\,1,\,2,\dots,
\eeqa
where $\bsy{a}^n\times\bsy{b}$ is the tensor defined by the components $\big(\bsy{a}^n\times\bsy{b}\big)_{i_1\dots i_n}=a_{i_1}\dots a_{i_{n-1}}(\bsy{a}\times\bsy{b})_{i_n}$.
Let us consider the process of {\bf STF} reducing for the multipole tensors starting with  the rank $N+1$  in the electric case, and  $N$ in the magnetic one. A general formula for the resulting tensors $\widetilde{\psft}^{(n)}$, for $n=1,\,\dots,\,N$ and  $\widetilde{\msft}^{(n)}$, for $n=1,\,\dots, \,N-1$ is given below. It includes  all compensating terms obtained in this process 
\beqa\label{Pgen}
&~&\widetilde{\psft}^{(n)}=\Tcal\big({\psft}^{(n)}\big)+\suml^{[(N-n)/2]}_{k=1}\frac{(-1)^k}{c^{2k}}\,\frac{\d^{2k-1}}{\d t^{2k-1}}\,\Tsft^{(n)}_k,\nonumber\\
&~&\Tsft^{(n)}_k=(-1)^kc^{2k}\,\Tcal\left(A^{(n)}_k\lasf^k\big(\dot{\psft}^{(n+2k)}\big)
+\suml^{k-1}_{l=0}B^{(n)}_{k-1,\,l}\lasf^l\,\Ncal^{2k-2l-1}\big(\msft^{(n+2k-1)}\big)\right),
\eeqa
and 
\beqa\label{Mgen}
\widetilde{\msft}^{(n)}=\Tcal\big(\msft^{(n)}\big)
+\Tcal\left(\suml^{[(N-n-1)/2]}_{k=1}\frac{\d^{2k}}{\d t^{2k}}\suml^k_{l=0}C^{(n)}_{kl}\lasf^l\Ncal^{2k-2l}\big(\msft^{(n+2k)}\big)\right).
\eeqa
The coefficients are given in a compact algebraic form in Ref. \cite{cvcs}: 
\beqa\label{ABC}
A^{(n)}_k&=&\frac{1}{2^k\,c^{2k}}\,\frac{n}{n+2k},\nonumber\\
B^{(n)}_{k,l}&=&\frac{(-1)^{k-l+1}\al}{2^l\,c^{2k+2}}\,\frac{n(n+2l)!}{(n+2k+1)(n+2k+1)!},\nonumber\\
C^{(n)}_{k,l}&=&\frac{(-1)^{k-l}}{2^l\,c^{2k}}\frac{n(n+2l)!}{(n+2k)(n+2k)!}.
\eeqa
The reader is encouraged to apply these formulae for the cases $N\ge 4$ and to find convenient intermediary  calculation. We point out that in equation \eref{Pgen}) the normalization of the quantities $\Tsft^{(n)}$ is chosen such that the quantities $\Tsft^{(n)}_1$ coincide with the electric toroidal moments given in literature at least for $n$ up to $3$.
We present  the results given in Ref. \cite{cvcs} concerning the total radiated power ${\mathcal I}$, the recoil force $\bsy{F}_r$ and angular momentum loss $\rmd \bsy{L}/\rmd t$:
\beqa\label{iepsmu}
&~&{\mathcal I}=\frac{\al^2}{4\pi\eps_0c^3}
\suml_{n\geq 1}\frac{n+1}{nn!(2n+1)!!c^{2n}}\left[\big(\msftr^{(n)}_{,\,n+1}\ct
\msftr^{(n)}_{,\,n+1}\big)+ 
\frac{c^2}{\al^2}\big(\psftr^{(n)}_{,\,n+1}\ct\psftr^{(n)}_{,\,n+1}\big)\right]\ ,
\eeqa

\beqa\label{Frf}
\bsy{F}_R&=&-\frac{\mu_0}{2\pi c^3}\suml_{n\geq 1}\frac{1}{c^{2n}}
\left\{
\frac{n+2}{(n+1)!(2n+3)!!}\big[\msftr^{(n)}_{,\,n+1}\ct\msftr^{(n+1)}_{,\,n+2}
+\frac{c^2}{\al^2}\big(\psftr^{(n)}_{,\,n+1}\ct\psftr^{(n+1)}_{,\,n+2}\big]\right.\nonumber\\
&&\left.-\frac{c^2}{\al} \frac{1}{n!n(2n+1)!!}\evec_i\eps_{ijk}
\big(\msftr^{(n)}_{,\,n+1}\ct n-1\ct\psftr^{(n)}_{,\,n+1}\big)_{jk}\right\}\ ,
\eeqa
with the obvious notation,
$$ \big(A^{(n)}\ct n-1\ct B^{(n)}\big)_{jk}=A_{i_1,\dots i_{n-1}j}
B_{i_1\dots i_{n-1}k}$$
and 
$$f_{,\,k}=\frac{\d^k}{\d t^k}f,$$

\beqa\label{finmomf}
\frac{\rmd\bsy{L}}{\rmd t} &=&\frac{\mu _{0}}{4\pi c}\sum\limits_{n\geq 1}%
\frac{1}{c^{2n}}\left\{- \frac{n+1}{n!(2n+1)!!}\vec{e}_{i}\varepsilon
_{ijk}\times \right.\nonumber \\
&&\times \left[ \frac{c^{2}}{\alpha ^{2}}\left( \psftr_{,n}^{\left( n\right)
}\ct n-1\ct \psftr_{,n+1}^{\left( n\right) }\right) _{jk}+\left(
\msftr_{,n}^{\left( n\right) }\ct n-1\ct \msftr_{,n+1}^{\left( n\right)
}\right) _{jk}\right] + \\
&&+\frac{n+2}{\al n!(2n+3)!!}\left[ \psftr_{,n+1}^{\left( n+1\right)
}\cdot \msftr_{,n+1}^{\left( n\right) }-\msftr_{,n+1}^{\left( n+1\right) }\cdot
\psftr_{,n+1}^{\left( n\right) }
 \left. +\msftr_{,n}^{\left( n\right) }\cdot \psftr_{,n+2}^{\left( n+1\right)
}-\psftr_{,n}^{\left( n\right) }\cdot \msftr_{,n+2}^{\left( n+1\right) }\right]
\right\}\ .\nonumber
\eeqa
%***************************

\end{document}